\documentclass[11pt]{amsart}
\usepackage{geometry}                
\usepackage{graphicx}
\usepackage{amssymb}
\title{Target Density Normalization for Markov Chain Monte Carlo Algorithms}
\author{Allen Caldwell, Chang Liu \\
Max Planck Institute for Physics, Munich, Germany}
\date{October 27, 2014}                                           

\begin{document}
\maketitle
\begin{abstract}
Techniques for evaluating the normalization integral of the target density for Markov Chain Monte Carlo algorithms are described and tested numerically. It is assumed that the Markov Chain algorithm has converged to the target distribution and produced a set of samples from the density.  These are used to evaluate sample mean, harmonic mean and Laplace algorithms for the calculation of the integral of the target density.  A clear preference for the sample mean algorithm applied to a reduced support region is found, and guidelines are given for implementation.
\end{abstract}

\section{Introduction}
Markov Chain Monte Carlo (MCMC) algorithms~\cite{ref:MCMC} are often used to generate samples distributed according to non-trivial densities in high dimensional spaces. Many algorithms have been developed that allow MCMCs to produce samples $\Lambda$ from an unnormalized target density $f(\lambda)$:
$$ \Lambda \sim f(\lambda)\; .$$

In many applications, it is desirable or even necessary to be able to normalize the target density.  I.e., to calculate

\begin{equation}
\label{eq:integral}
 I  \equiv  \int_{\Omega}  f(\lambda) d\lambda 
 \end{equation}
where $\Omega \in \mathbb{R}^D$ is the support of $f$.  This integral can be computationally very costly or impossible to perform with standard techniques if the volume where the target $f$ is non-negligible occupies a very small part of the total volume of $\Omega$.  

An important area where such integration is necessary is for Bayesian data analysis~\cite{ref:Jeffreys,ref:Jaynes}.  Bayes' formula reads, for a given model $M$,
\begin{equation}
P(\lambda | {\rm Data},M) = \frac{ P({\rm Data}| \lambda,M) P_0(\lambda|M) }{P({\rm Data}|M)}
\end{equation}
where here $\lambda$ are the parameters of the model and the data are used to extract probabilities for possible values of $\lambda$.  The denominator is usually expanded using the Law of Total Probability and written in the form 
\begin{equation}
Z=P({\rm Data}|M)=\int P({\rm Data}|\lambda,M) P_0(\lambda|M) d\lambda 
\end{equation}
and goes by the names `evidence', or `marginal likelihood', and is the type of integral that we want to be able to calculate (here the data are fixed and $f(\lambda) = P({\rm Data}|\lambda,M) P_0(\lambda|M) $). An example use of $Z$ is for the calculation of Bayes Factors in the comparison of two models:
$$
{\rm BF} \equiv \frac{P({\rm Data}| M_A)}{P({\rm Data}| M_{B})} = \frac{Z_A}{Z_{B}} \; .
$$
Another application where the calculation of a normalization can be very important is in the parallelization of the MCMC algorithm~\cite{ref:parallel}.  While the MCMC approach has very attractive features, it is often slow in its execution due to the nature of the algorithm.  A goal is therefore to parallelize the computations needed to map out the target density.  This looks at first sight difficult since the MCMC algorithms are by construction serial.  A parallelization of the calculations can however be achieved via a partitioning of the support.  I.e., we partition $\Omega$ into sub volumes $\omega_i$ with 
$$\cup \omega_i = \Omega \hskip 2cm \omega_i \cap \omega_j = \emptyset \; {\rm for} \; i\neq j $$
and we run a separate MCMC sampling for each sub volume $\omega_i$.  In order to have a final set of samples representing the target density over the full support, we need to know the relative probabilities for the different sub volumes.  I.e, we need
$$ I_i \equiv  \int_{\omega_i}  f(\lambda) d\lambda \hskip 2cm \left( \sum_i I_i = I \right)$$
The samples in the different regions are then given weights $ \propto \frac{I_iN}{IN_i}$
with $N_i$ the number of samples from $f(\lambda)$ in $\omega_i$ and $\sum_i N_i \equiv  N$.

\section{Techniques}

A variety of techniques to calculate the evidence in Bayesian Calculations have been successfully developed.  A summary can be found in~\cite{ref:evidence}, where a number of MCMC related techniques are reviewed, including Laplace's method~\cite{ref:Laplace}, harmonic mean estimation~\cite{ref:HME}, Chib's method~\cite{ref:Chib}, annealed importance sampling techniques~\cite{ref:Robert}, Nested Sampling~\cite{ref:Skilling} and thermodynamic integration methods~\cite{ref:thermo,ref:Friel}.  

We are here specifically interested in testing techniques directly applicable in an MCMC setting, and which is independent of the specific MCMC algorithm.  We assume that the MCMC algorithm has been successfully run to extract samples according to the target density, and the goal is to provide an algorithm for calculating the normalization (or evidence).  Given our requirements, only arithmetic mean estimation (AME), harmonic mean estimation (HME) and Laplace methods are directly applicable.  Using AME and HME methods directly is known to fail in many situations, and the Laplace method is only applicable if the target density is Gaussian.  We introduce the use of a reduced integration volume and normalization using the MCMC output to improve the AME and HME performance.  After a description of the techniques, we report on numerical investigations of the different approaches using samples from the MCMC code BAT~\cite{ref:BAT}.

\subsection{Reduced Volume Evaluation}
Assuming the MCMC has been successfully run to extract samples according to $f(\lambda) $, one of the quantities directly retrievable from the MCMC output is an estimate of the parameter values at the global mode: $\Lambda^*$ is in the neighborhood of $\lambda^*$.  I.e., we know approximately where the integrand in Eq.~\ref{eq:integral} has its maximum.

We note that
\begin{equation}
\label{eq:scheme}
r \equiv  \frac{\int_{\omega} f(\lambda) d\lambda}{I} \approx \frac{N_{\Delta}}{N_{\rm MCMC}}\equiv \hat{r}
\end{equation}
with $\omega$ a sub support of $\Omega$ is directly estimated from the MCMC output by counting the fraction of samples falling within $\omega $, $N_{\Delta}$ (the reason for this notation will become clear below).
I.e., the task of evaluating $I$ reduces to integrating the function $f(\lambda)$ over a well-chosen region - presumably a small region around $\lambda^*$ and dividing by $\hat{r}$.  This integral can be much simpler to evaluate than the integral over the full support.

%
%

\subsection{Choice of Region $\omega$}
In the following, we use a simple hypercube for our integration region. From the MCMC samples, we can construct the marginalized distributions along each of the $\lambda$ dimensions.   We define an interval along each dimension centered at $\Lambda^*$ with width which is a multiple of the standard deviation (we use the symbol $\Delta$ to represent this factor).   The optimum value of $\Delta$ depends on the dimensionality of the problem as described below.  Another option would be to produce a covariance matrix of the $\Lambda$ for sampling using a multivariate normal distribution if desired, but this was not found necessary in the examples we have studied.

\subsection{Arithmetic Mean Estimation}
\label{sec:SM}

The integral in the numerator in Eq.~\ref{eq:scheme} can presumably be determined in a straightforward way since now we are focusing on a small volume with significant mass.  The standard importance sampling approximation is given by

\begin{eqnarray*}
I_{\Delta} &\equiv& \int_{\omega} f(\lambda) d\lambda \\
  &=& \int_{\omega} \frac{f(\lambda)}{g(\lambda)}  g(\lambda)d\lambda \\
  &\approx& \frac{1}{N_{\rm SM}} \sum_{\Lambda \in \omega} \frac{f(\Lambda)}{g(\Lambda)}|_{\Lambda\sim {g(\lambda)}}
\end{eqnarray*}
where our sampling probability density is given by $g(\lambda)$.  $N_{\rm SM}$ is the number of samples used in the  calculation. If we choose for $g(\lambda)$ a uniform distribution in the hypercube, then we have the well-known sample mean result
\begin{equation}
\label{eq:SM}
I_{\Delta} \approx \frac{V_{\Delta}}{N_{\rm SM}} \sum_{\Lambda \in \omega}  f(\Lambda) \equiv  \hat{I}_{\Delta} 
\end{equation}
with $V_{\Delta}$ the volume of the hypercube.  
%
%
Our estimator for $I$ is then
\begin{equation}
\hat{I}_{AME} \equiv  \frac{\hat{I}_{\Delta}}{\hat{r}} \;\; .
\end{equation}

We will use this simplest version of the estimator for our examples below.

\subsection{Uncertainty Estimate}

Assuming unbiased Gaussian distributions for $\hat{I}_{\Delta}$ and $\hat{r}$ about their true values, we can estimate the uncertainty for $I$ with 
\begin{equation}
\label{eq:uncertainty}
\sigma_I = \sqrt{ \left( \frac{\sigma_{r} \cdot \hat{I}_{\Delta}}{\hat{r}^2} \right)^2 + \left( \frac{\sigma_{I_{\Delta}}}{\hat{r}} \right)^2 }
\end{equation}
where
$$\sigma_{r}=\sqrt{\frac{\hat{r}(1-\hat{r})}{N_{\rm ESS}}} \;\; .$$
The effective sample size~\cite{ref:ESS}  is defined here as
\begin{equation}
\label{eq:ESS}
N_{\rm ESS} \equiv \frac{N_{\rm MCMC}}{1 + 2\sum_{j=1}^{N_{\rm MCMC}} \rho_j}
\end{equation}
with  the autocorrelation function at $j$ defined for our MCMC sample as
\begin{eqnarray}
\rho_j &\equiv& \frac{\sum_{i=1}^{D} (\Lambda^j_i \Lambda^{j+1}_i - <{\Lambda_i}>^2)}{\sum_{i=1}^D \sigma_i^2} \\
<{\Lambda_i}> &=& \frac{1}{N_{\rm MCMC}} \sum_{j=1}^{N_{\rm MCMC}} \Lambda_i^j \\
\sigma_i^2 &=&  \frac{1}{N_{\rm MCMC}} \sum_{j=1}^{N_{\rm MCMC}} (\Lambda_i^j)^2 -  <{\Lambda_i}>^2\; .
\end{eqnarray}
In these equations, the subscript $i=1 \ldots D$ labels the component of $\Lambda$, while the index $j$ labels the iteration in the MCMC.

The uncertainty from the sample mean integration is estimated by separating the sample mean calculation of $\hat{I}_{\Delta}$ into $K$ batches and looking at the variance of these calculations:
\begin{equation}
\sigma_{I_{\Delta}} = \sqrt{\frac{\sum_{k=1}^{K} \left(\hat{I}_{\Delta,k} - <\hat{I}_{\Delta}  >\right)^2}{K(K-1)} } \; .
\end{equation}


With these definitions, we are able to report both an estimate for our integral and an uncertainty.  These will be compared to accurately calculated values for the chosen examples in the following sections.

\subsection{Harmonic Mean Estimation}
\label{sec:HME}
The HME~\cite{ref:HME} value for $I$ can be calculated as follows:
\begin{eqnarray}
E\left[ \frac{1}{f(\lambda)}\right]_{\hat{f}(\lambda)} &=& \int_{\Omega} \frac{1}{f(\lambda)} \cdot \frac{f(\lambda)}{I} d\lambda \\
&=& \frac{V}{I} 
\end{eqnarray}
where $\hat{f}(\lambda)$ is the normalised target density and $V$ is the total volume of the support.  The HME estimator is then
\begin{equation}
\hat{I} = \frac{N_{\rm MCMC}V}{\sum_{\Lambda \in \Omega} \frac{1}{f(\Lambda)}} \; .
\end{equation}

This calculation is performed directly from the MCMC output from which the samples $\Lambda$ as well as $f(\Lambda)$ are available, and does not require an extra sample mean calculation as in the AME scheme.  However, it can be unstable because of samples occurring (or missing) in regions where $f(\lambda)$ is small (relative to other regions).  We can improve the estimation, as originally noted in~\cite{ref:GD}, by limiting ourselves to a small volume around the mode.  Using the same notation as above, we can write
\begin{equation}
\label{eq:HME}
\hat{I}_{\rm HME} \equiv \frac{N_{\rm MCMC}V_{\Delta}}{\sum_{\Lambda \in \omega} \frac{1}{f(\Lambda)}} \; .
\end{equation}
where now only the samples in the restricted support $\omega$ are used.   The uncertainty in the estimate is calculated by separating the MCMC samples included in our integration region into batches and looking at the variation of these estimates. 

\subsection{Laplace Method}
In this approach, the target distribution is assumed to be represented by a (multivariate) Gaussian distribution.
The estimator for the normalization is then
\begin{equation}
\hat{I}_{\rm L} \equiv (2\pi)^{D/2} |\Sigma^*|^{1/2} f(\Lambda^*)
\end{equation}
where the target density is evaluated at the mode returned from the MCMC and $|\Sigma^*|$ is the determinant of the covariance matrix evaluated numerically from the samples $\Lambda$.  This method is clearly only expected to work in cases where the assumption of normality is valid.

\section{Examples}
\subsection{Product of one-dimensional Gaussians}
We start with a simple example - the target density is the product of a number of Gauss functions depending on only one parameter - to describe our  testing procedures in detail.  We then move on to more complicated examples in multivariate spaces, including functions with degenerate modes.  All MCMC calculations were performed using the BAT program, with samples from the target density taken after convergence of the MCMC algorithm.

We start with the following target function:
\begin{equation}
f(\mu) = \prod_{i=1}^{m} \frac{1}{\sqrt{2\pi}\sigma}e^{-\frac{1}{2}\left(\frac{x_i-\mu}{\sigma}\right)^2} \;\; .
\end{equation}
This type of function could, e.g., be the likelihood function constructed for producing an estimate of a quantity, $\mu$, given $m$ measurements, $\{x_i\}$, with a sampling distribution modeled by a Gaussian probability distribution of fixed width $\sigma$. The normalization integral for the target can be performed analytically assuming the volume of interest extends well beyond  the extreme values of the $x_i$.  For the more general case of the product of $m$ $D$-dimensional uncorrelated Gaussian functions with known variances $\sigma_j^2, \; (j=1,\ldots D)$, the integral is given by 

\begin{eqnarray*}
 I&=& \int_{\Omega} f(\vec{\mu}) d\vec{\mu} \\
 &=& \int_{\Omega} \left[ \prod_{i=1}^{m} \frac{1}{(2\pi)^{D/2}|\Sigma|^{1/2}}e^{-\frac{1}{2}(\vec{x}_i-\vec{\mu})^T\Sigma^{-1} (\vec{x}_i-\vec{\mu})} \right] d\vec{\mu} \\
 &=&\frac{1}{(2\pi)^{D\cdot (m-1)/2}\vert\Sigma\vert^{(m-1)/2} m^{D/2}} \exp\left(-\sum_{j=1}^D \frac{{\rm Var}[x_j]}{2\sigma_j^2} \right) 
\end{eqnarray*}
where $\Sigma$ is the (diagonal) covariance matrix.

For our concrete example, we take $m=100$ and generate random values of $x$ from a Gauss distribution of mean zero and unit standard deviation, and we find for the generated values  $\ln I_{\rm true} = -67.18$.  In evaluating the integral, we take for the support
$\Omega =\{ -100 \leq \mu \leq 100\}$.

\begin{figure}[htbp] 
   \centering
   \includegraphics[width=7cm]{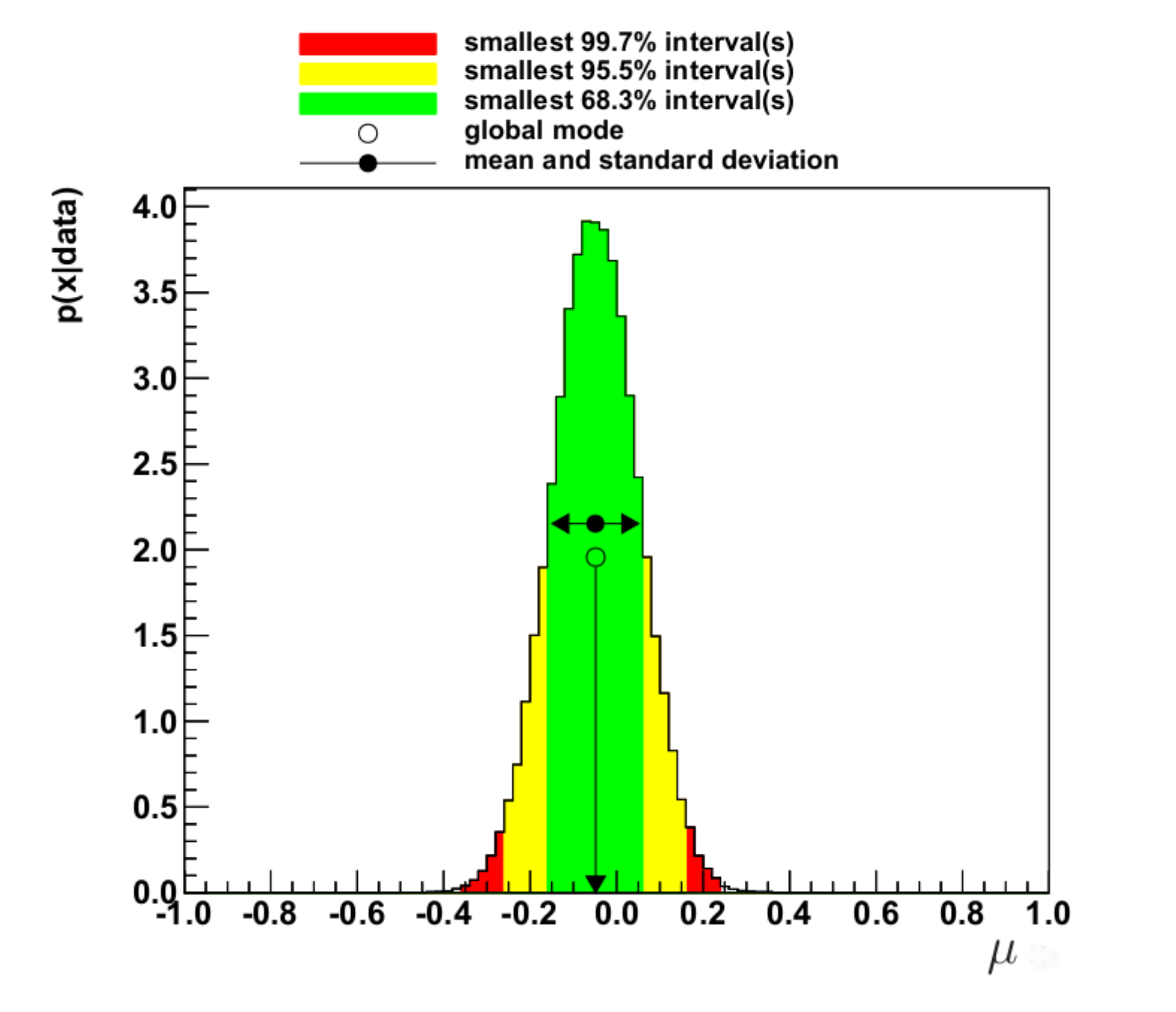}    \includegraphics[width=7cm]{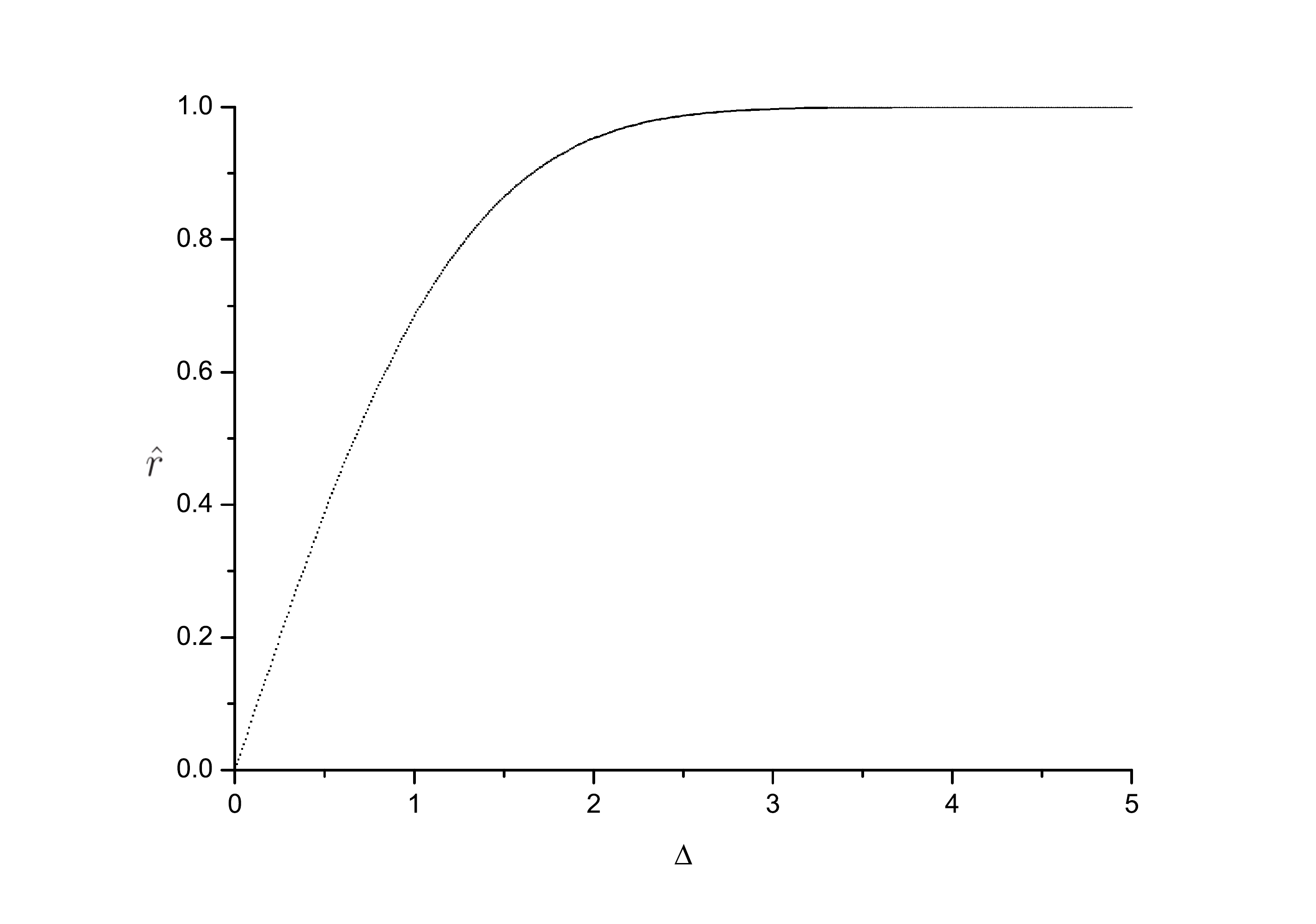} 
   \caption{Left) Distribution of samples from the MCMC algorithm BAT~\cite{ref:BAT} for the product of 100 Gauss distributions as described in the text.   Right) Fraction of MCMC samples falling within the interval of length $2\Delta$ as a function of the value of $\Delta$ (in units of the standard deviation of the distribution).}
   \label{fig:MCMC1D}
\end{figure}

We then use $N_{\rm MCMC}=10^5$ samples from the MCMC output to find an estimate for the mode of $f(\mu)$ and to calculate the standard deviation for $\mu$. The distribution of samples from $f(\mu)$ from the MCMC are displayed in Fig.~\ref{fig:MCMC1D}(left).  The mode of the samples is found at $\mu^*=-0.05$ and the standard deviation is found to be $\sigma_{\mu}=0.10$.  The effective sample size for this set of samples is $N_{\rm ESS}=3.9\cdot 10^4$.

 The dependence of $\hat{r}$ on the chosen value of $\Delta$ is also shown in Fig.~\ref{fig:MCMC1D}(right) for 500 values of $\Delta$ ranging from $0.01$ to $5$ in steps of $0.01$.  For a one-dimensional Gaussian target density, which is what we have here, the expectation is that 68~\%  of MCMC samples occur within $\Delta=1$ and 95~\% occur within $\Delta=2$, and this is indeed what is found.

\begin{figure}[htbp] 
   \centering
   \includegraphics[width=7cm]{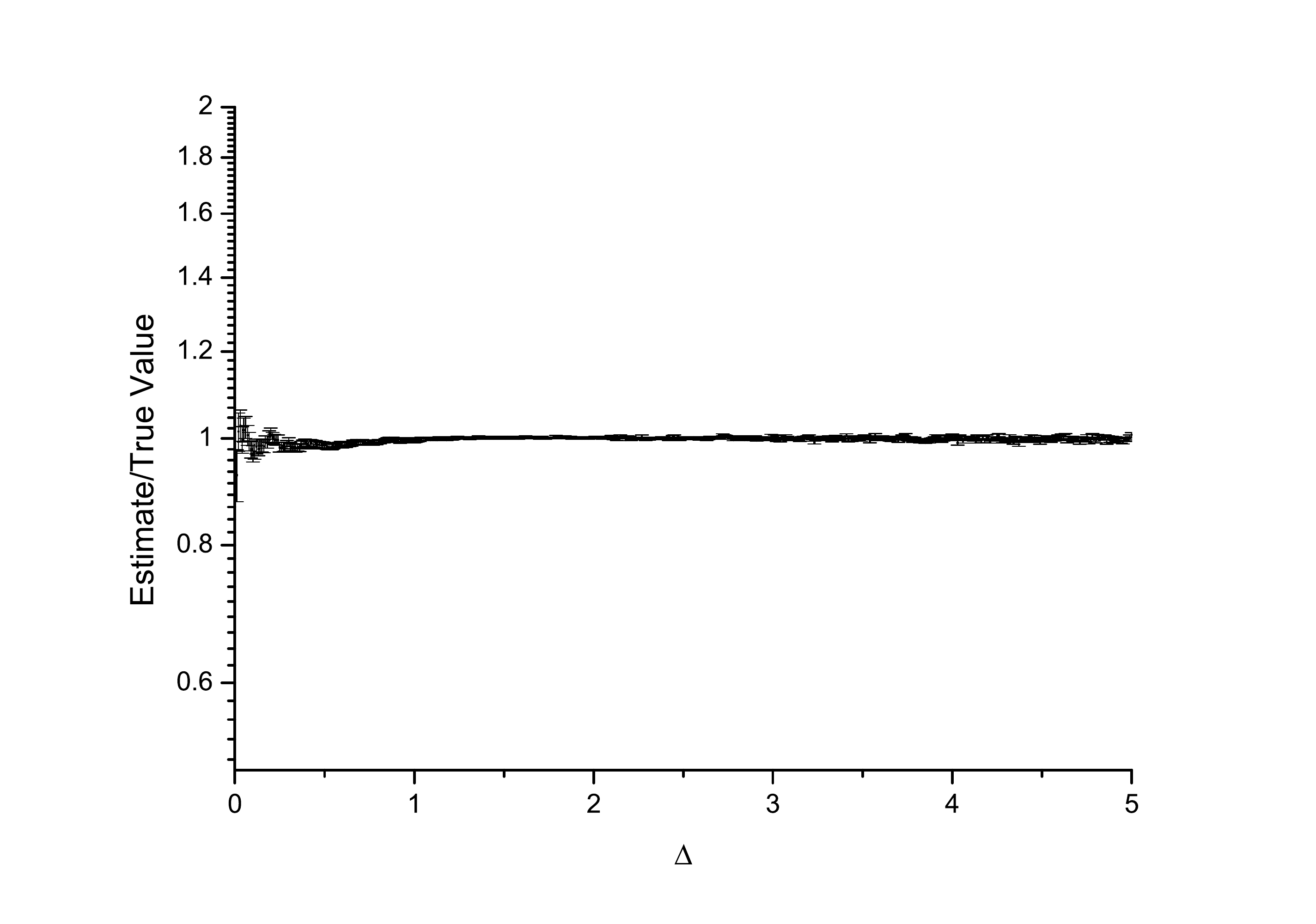}    \includegraphics[width=7cm]{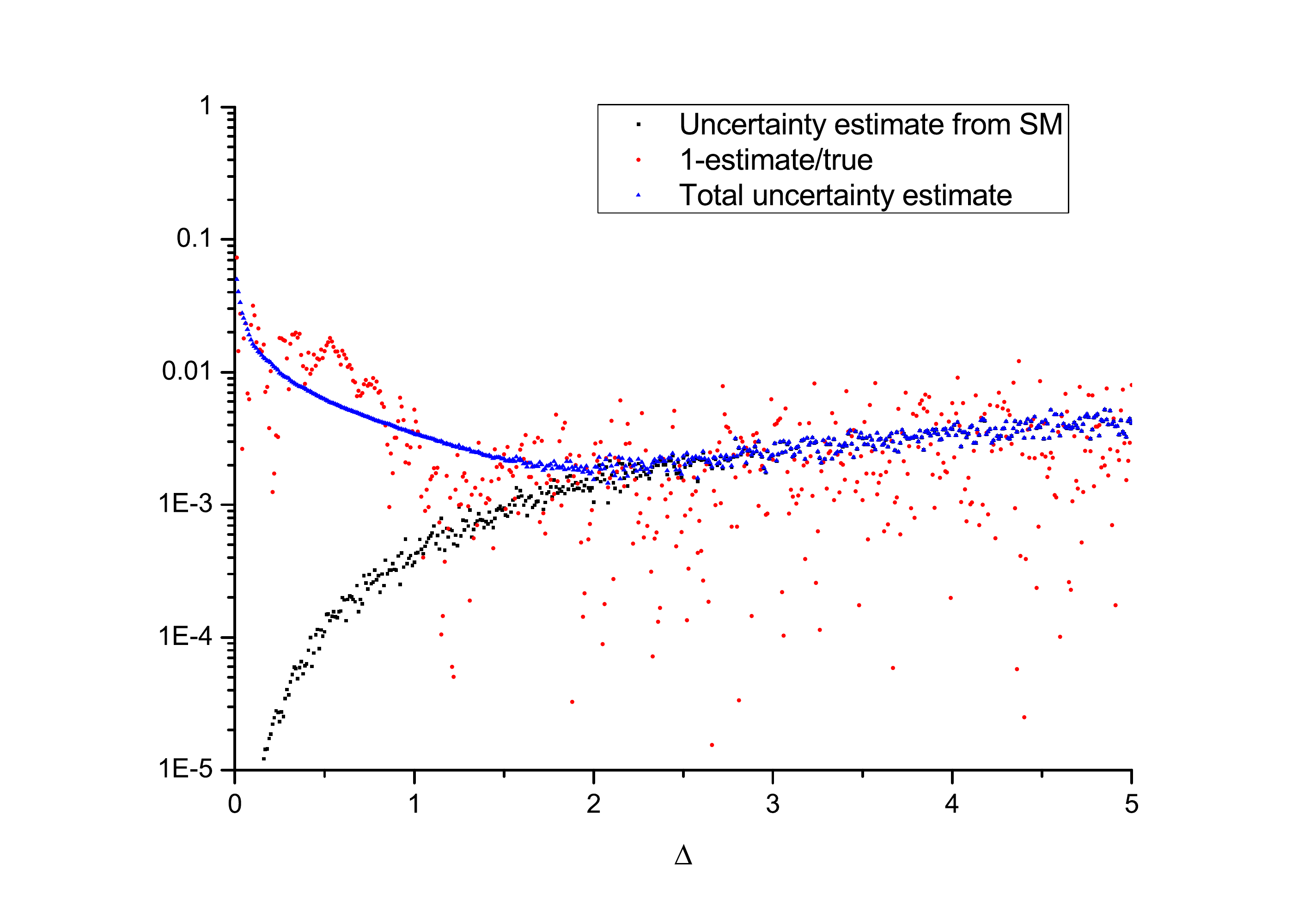} 
   \caption{ Arithmetic Mean results for the one-dimensional Gauss product example. Left) $\hat{I}_{\rm AME}$ as a function of $\Delta$, scaled by the true value.  The error bars correspond to the estimated uncertainty.  Right)  The actual error $\vert \hat{I}_{\rm AME}/I_{\rm true} -1\vert$ (red), the estimated uncertainty $\sigma_{I_{\Delta}}/I_{\rm true}$  (black) and the total estimated uncertainty $\sigma_I/I_{\rm true}$ (blue) as a function of $\Delta$. }
   \label{fig:evidence1D}
\end{figure}

We then perform a sample mean calculation with $N_{\rm SM}=10^5$ samples for each of the different choices of $\Delta$.  For each calculation, we extract a value of $\hat{I}_{\rm AME}$ as described in section~\ref{sec:SM} as well as an estimate of the uncertainty.  The extracted values of  $\hat{I}_{\rm AME}$  (divided by the true value) are shown as a function of $\Delta$ in Fig.~\ref{fig:evidence1D}(left).  The error bars are the estimated one standard deviation uncertainties.  We observe  small systematic deviations of the results for small values of $\Delta$ resulting from the inaccurate determination of $r$ from the MCMC samples (note that the MCMC was only run once, so that the $\hat{r}$ values are correlated).  

To study the uncertainty estimation, we compare $\sigma_I/I_{\rm true}$ to $(\hat{I}_{\rm AME}-I_{\rm true})/I_{\rm true}$ at each value of $\Delta$.  The results are shown in Fig.~\ref{fig:evidence1D}(right).  In this figure, the red points indicate the absolute  value of $(\hat{I}_{\rm AME}-I_{\rm true})/I_{\rm true}$, the black points the estimated uncertainty coming from the sample mean calculation, $\sigma_{I_\Delta}/I_{\rm true}$, and the blue points the total estimated uncertainty, $\sigma_I/I_{\rm true}$.  We observe that our estimated uncertainty is accurate, and that there is a minimum of the uncertainty around $\Delta=2$.  The location of the minimum clearly depends on the number of samples chosen for the MCMC and sample mean calculations, but it is important that we can accurately estimate the uncertainty.  In this case, the arithmetic mean calculation is quite accurate even at large values of $\Delta$ since we are only working in one-dimension.

\begin{figure}[htbp] 
   \centering
   \includegraphics[width=7cm]{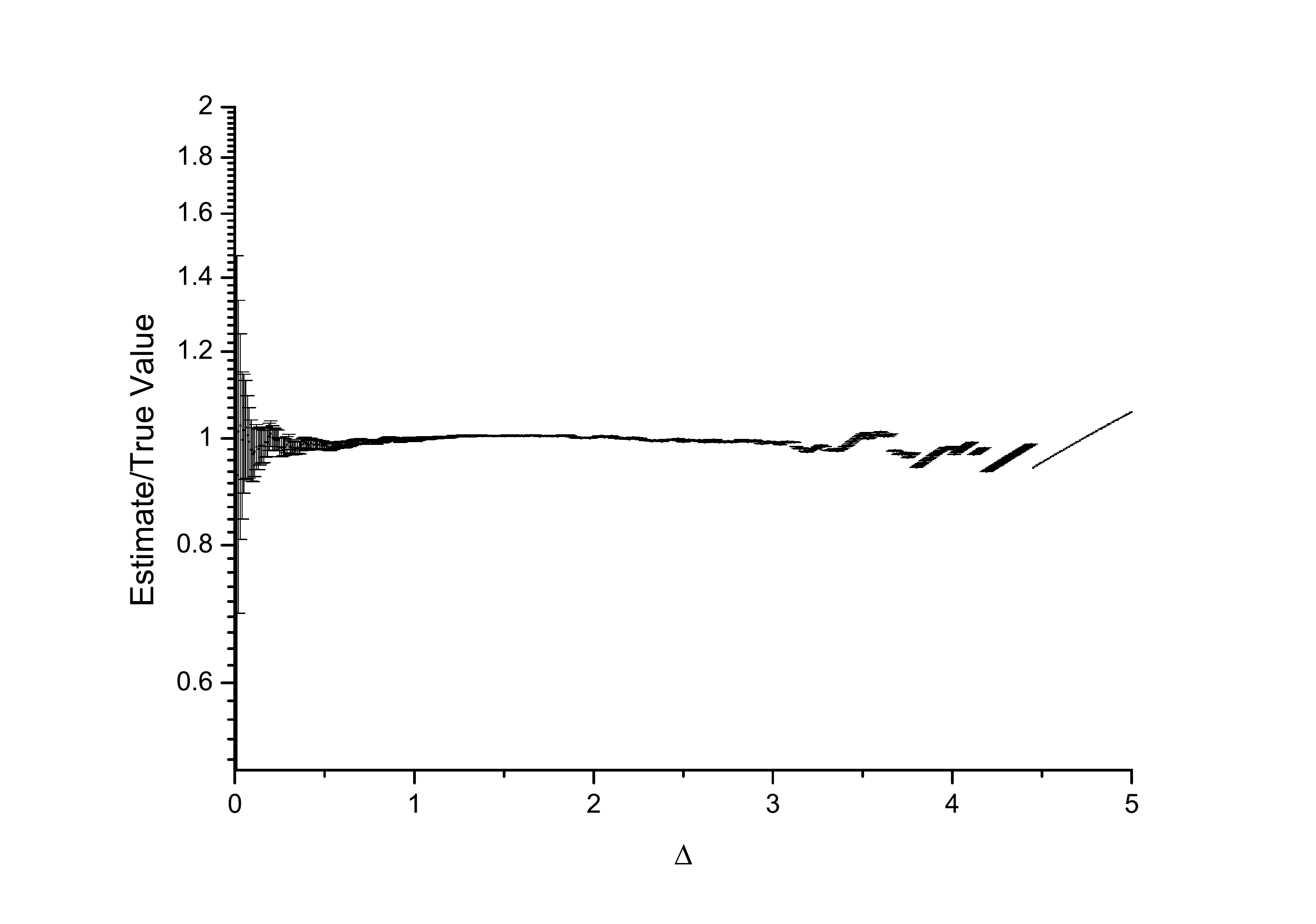}    \includegraphics[width=7cm]{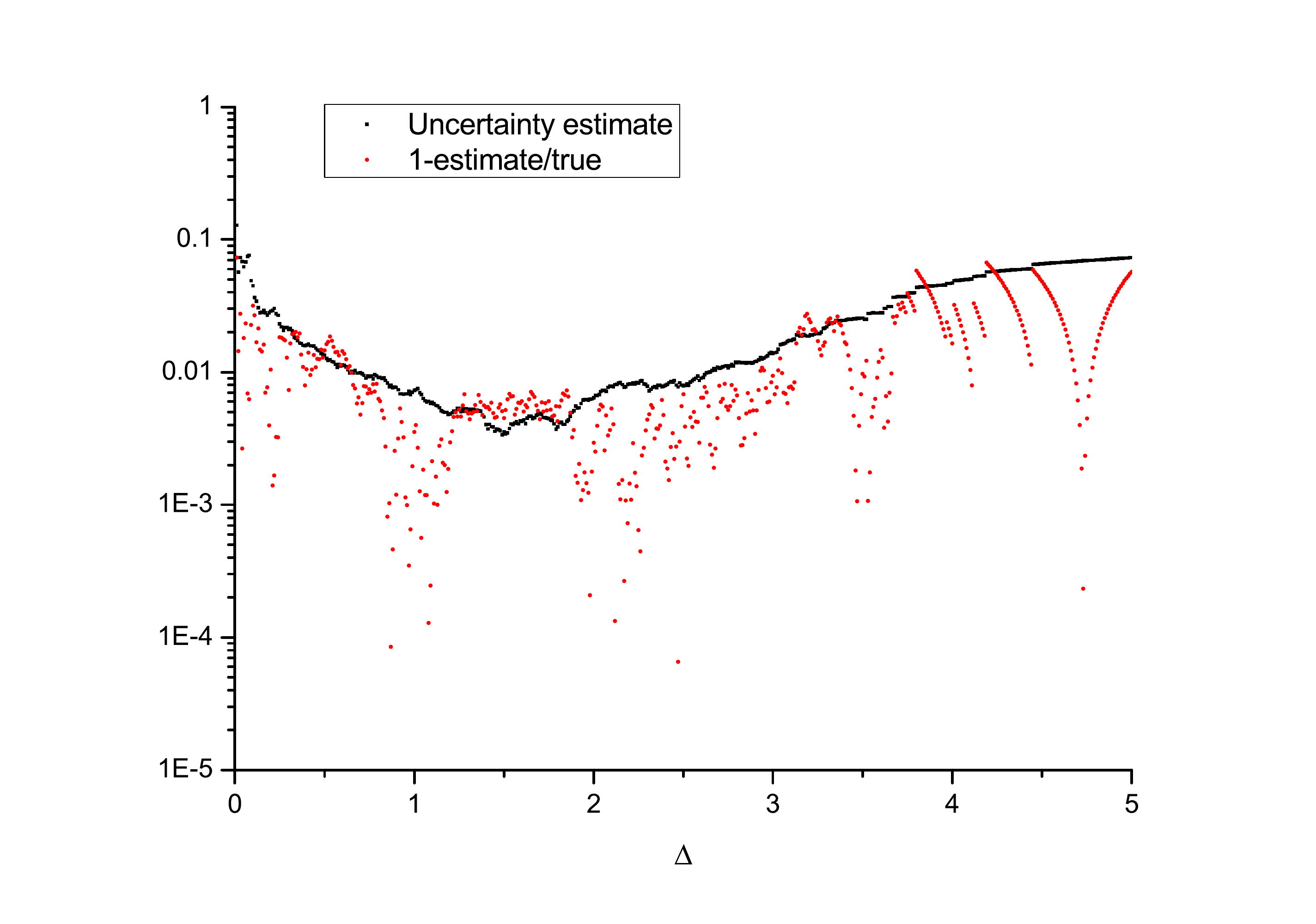} 
   \caption{Harmonic Mean results for the one-dimensional Gauss product example. Left)  $\hat{I}_{\rm HME}$ scaled by the true value as a function of $\Delta$.  The error bars correspond to the estimated uncertainty.  Right) The actual error $\vert \hat{I}_{\rm HME}/I_{\rm true} -1\vert $ and the estimated uncertainty as a function of $\Delta$. }
   \label{fig:HME}
\end{figure}

We now evaluate the Harmonic Mean estimate for $I$ as described in section~\ref{sec:HME}.  The estimate $\hat{I}_{\rm HME}$ as well as the absolute deviation from $I_{\rm true}$ as a function of $\Delta$ are shown in Fig.~\ref{fig:HME}.  We see for this example that the HME technique works well, and that accuracies of a fraction of 1~\% are possible from the HME estimation at $\Delta \approx 1.5$.  As $\Delta$ is increased, the HME estimation worsens since, although more of the MCMC samples are included, reducing the binomial uncertainty on $N_{\Delta}$, imperfect  sampling in the tails of the distribution plays a large role and we see the importance of limiting the range of the integration region for the HME calculation already with this simple one-dimensional example.  The uncertainty is somewhat worse than what was found for the AME calculation, but probably adequate for the majority of applications.  Also, the calculation did not require the extra step of performing a sample mean calculation.

As seen in Fig.~\ref{fig:MCMC1D}, the target density is Gaussian and therefore the Laplace method is expected to work well.  Indeed, the Laplace method yields an estimate within $0.2$~\%  of the true value in this example:  $\hat{I}_{\rm L}/I_{\rm true}=1.0016$.

\subsection{Product of Multivariate Gaussians}
We now move to a target density composed of a product of ten dimensional Gaussian distributions with non-diagonal covariance matrix.  The target function in this case is:
\begin{equation}
f(\vec{\mu}) = \prod_{i=1}^{m} \frac{1}{(2\pi)^{5}|\Sigma|^{1/2}}e^{-\frac{1}{2}(\vec{x}_i-\vec{\mu})^T\Sigma^{-1} (\vec{x}_i-\vec{\mu})} 
\end{equation}
where $\Sigma$ is the covariance matrix, assumed to be known,  and $m=20$.  The target function is ten-dimensional and has significant correlations among the ten parameters.  The values of $\vec{x}_i$ were chosen by generating random vectors using $\vec{\mu}_{\rm true}=\vec{0}$ and the following covariance matrix

$${\small
\Sigma=\begin{bmatrix}
1.0&  1.0 &  0.30&  0.43&  -0.14&  -0.86&  -0.22&  -0.84&  0.83&  -2.5\\
1.0&  2.0&  -0.14&  0.36&  0.14&  -0.08&  -0.45&  0.71&  0.07& -1.9\\
0.30&  -0.14&  3.0&  -0.04&  0.76&  0.85&  1.7&  -0.41&  1.2&  -0.78\\ 
0.43&  0.36& -0.04&  4.0&  -0.81&  -0.86&  -1.5&  -2.1&  0.17&  0.63\\ 
-0.14&  0.14&  0.76&  -0.81&  5.0&  2.5&  1.6& 1.5& 1.9&  0.46\\ 
-0.86&  -0.08&  0.85&  -0.86&  2.5&  6.0&  1.9&  4.1&  -0.28&  2.7\\ 
-0.22&  -0.45&  1.7&  -1.5&  1.6&  1.9&  7.0&  0.70&  1.4&  2.6\\ 
-0.84&  0.71&  -0.41&  -2.1&  1.5&  4.1&  0.70&  8.0&  -0.87&  2.5\\ 
0.83& 0.07&  1.2&  0.17&  1.9&  -0.28&  1.4&  -0.87&  9.0&  -4.0\\ 
-2.5&  -1.9&  -0.78&  0.63&  0.46&  2.7&  2.6&  2.5&  -4.0&  10\\ 

\end{bmatrix}
}
$$ 
and again could represent a type of situation found in a data analysis setting.  The integration region for $I$ was taken as a $10$D hypercube of side length $200$ centered on $\vec{\mu}=\vec{0}$.

The value for $I_{\rm true}$ can again be evaluated analytically by finding the similarity transformation that diagonalizes the covariance matrix. The expression of the integral in this case is 

\begin{equation}
 I=\frac{A}{(2\pi)^{N\cdot (m-1)/2}\vert\Sigma\vert^{(m-1)/2} m^{N/2}} \exp\left(-\sum_{j=1}^N {\rm Var}[x_j'] \right) .
\end{equation}
where $\vec{x}'=\sqrt{S} V\vec{x}$ and 
$\Sigma^{-1} =  V^{-1} S V $
with $S$ a diagonal matrix.

\begin{figure}[htbp] 
   \centering
  \includegraphics[width=9cm]{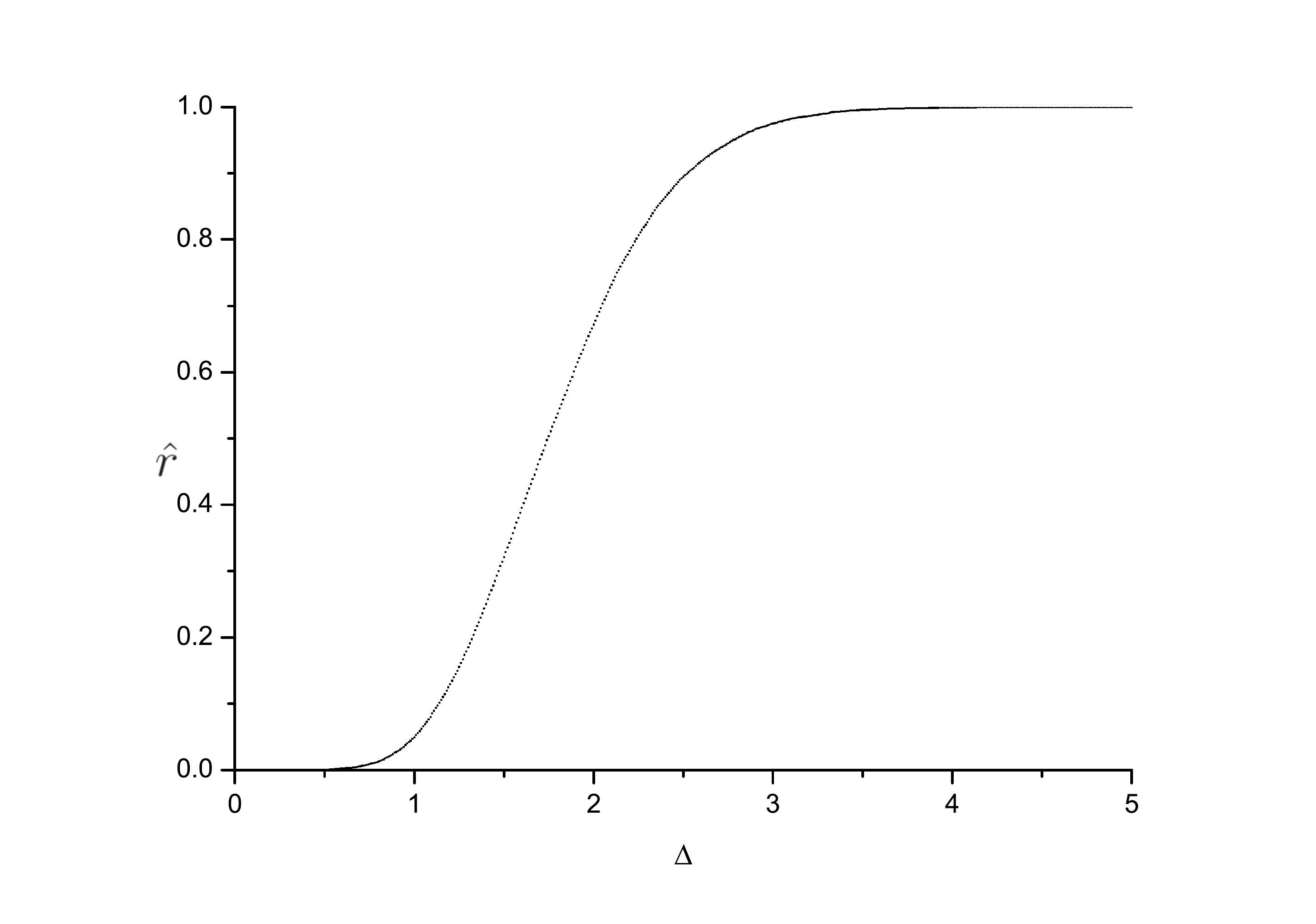} 
   \caption{Fraction of MCMC samples falling within the hypercube centered on the observed mode and of side length $2\Delta$ as a function of the value of $\Delta$ (in units of the standard deviation of the marginalized distribution) for the product of ten-dimensional correlated Gauss functions.}
   \label{fig:10Destimator}
\end{figure}

The true value of the integral for randomly generated data was evaluated using this expression and yielded
$\ln(I_{\rm true}) =-427.5$. The MCMC program BAT was used to sample from the target density with $5\cdot 10^5$ samples stored post-convergence (yielding  $N_{\rm ESS}=1.9 \cdot 10^5$).  The value of $\hat{r}$ is given as a function of $\Delta$ in Fig.~\ref{fig:10Destimator}.

\begin{figure}[htbp] 
   \centering
   \includegraphics[width=7cm]{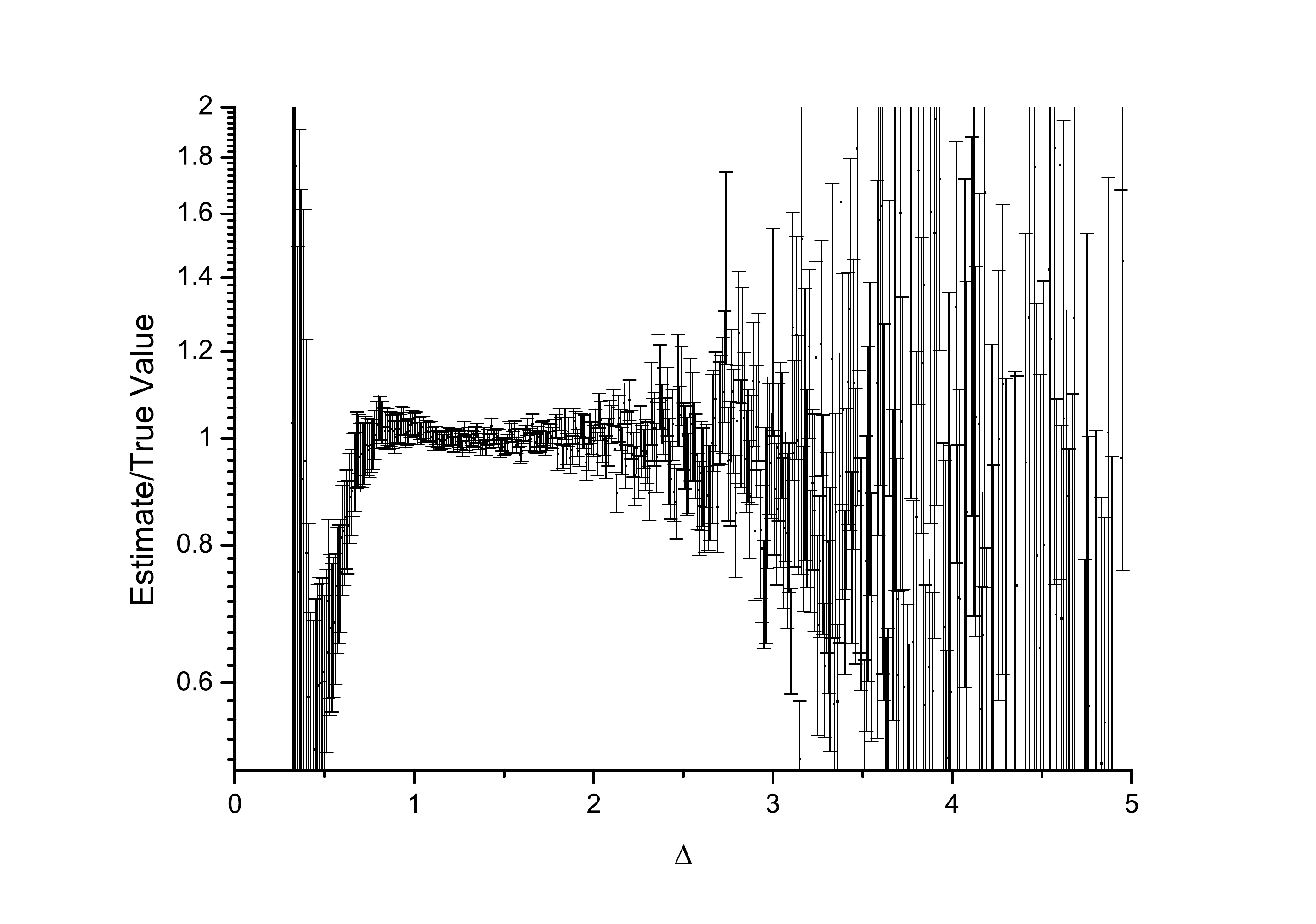}    \includegraphics[width=7cm]{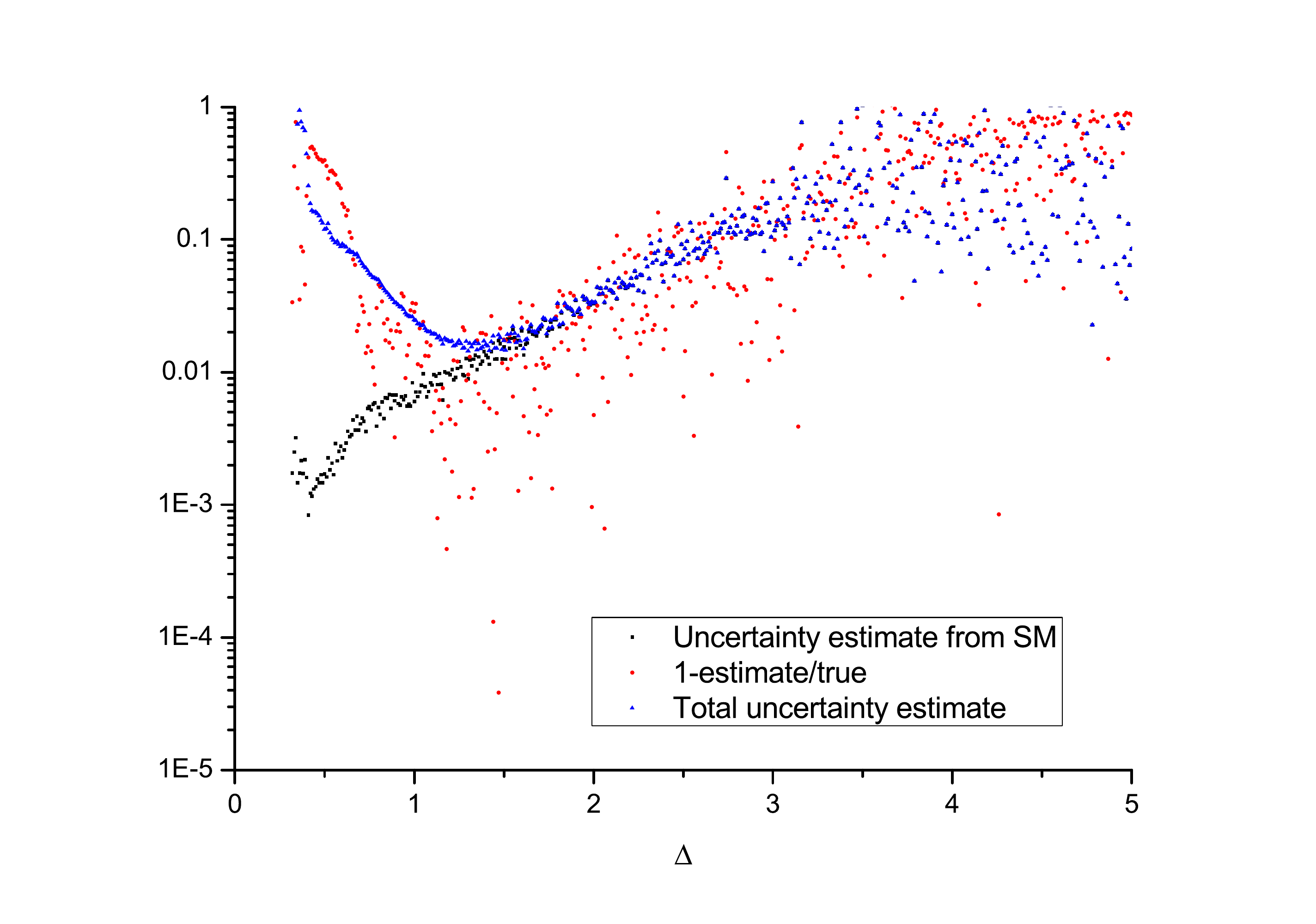} 
   \includegraphics[width=7cm]{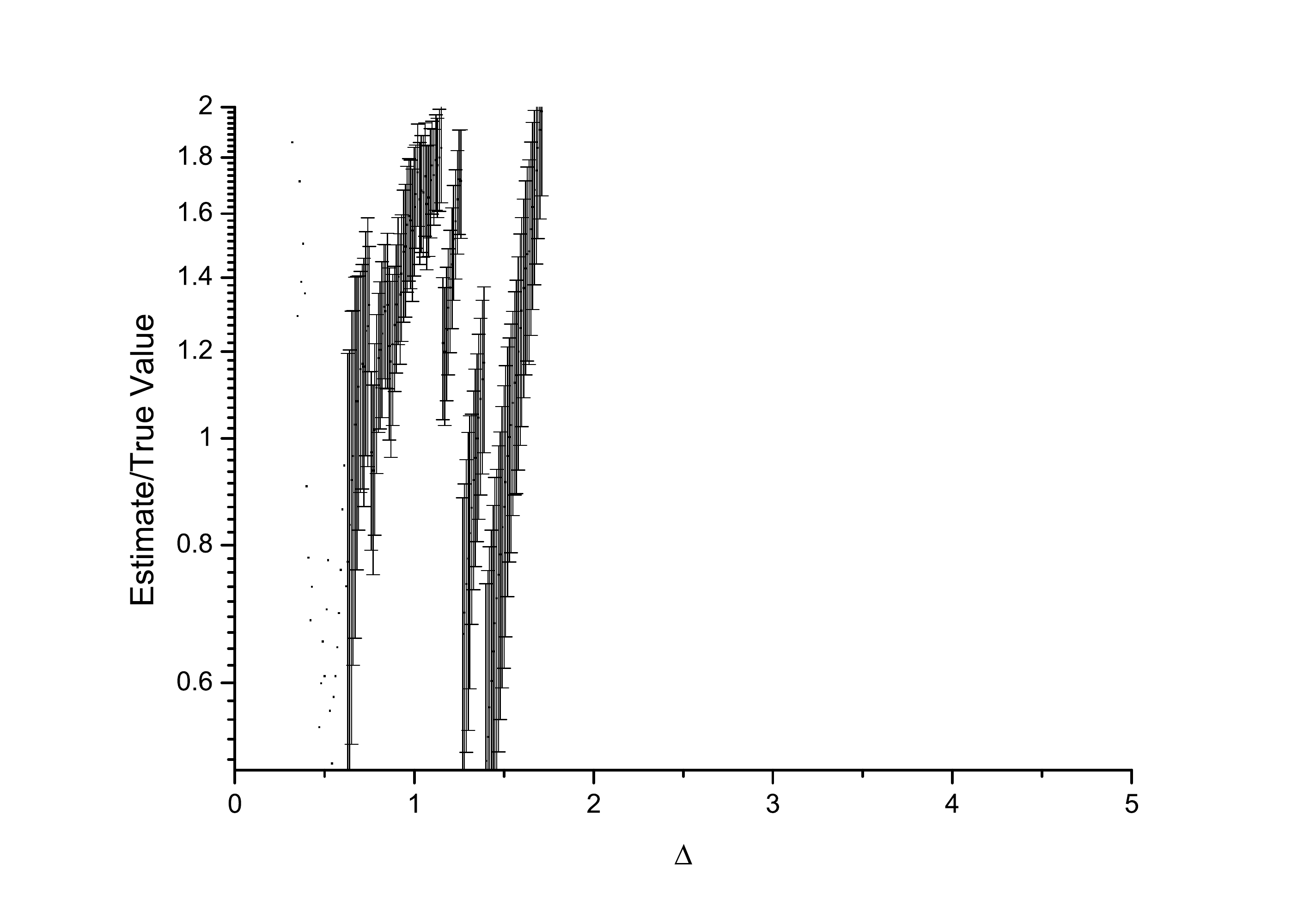}    \includegraphics[width=7cm]{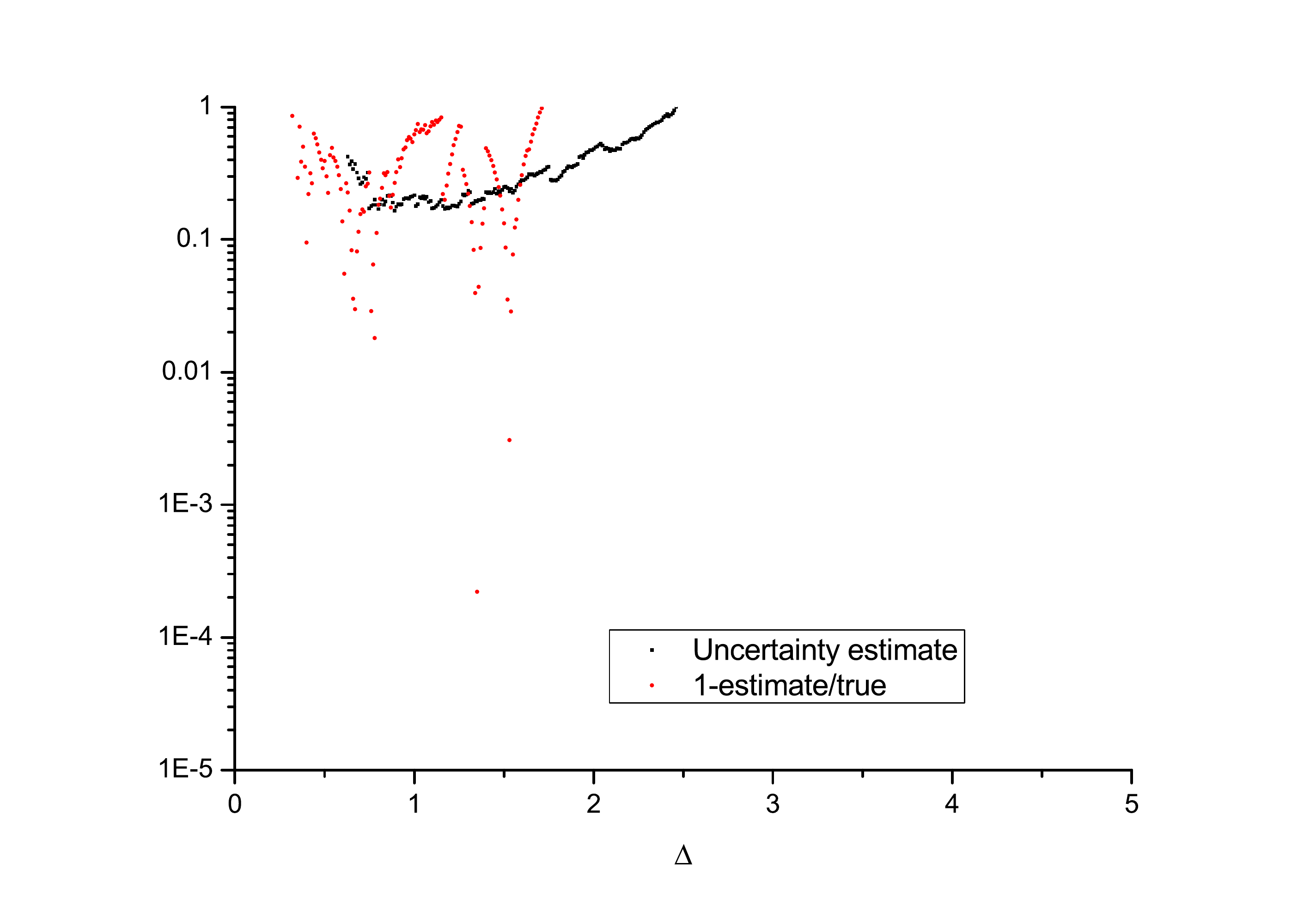} 
   \caption{10D correlated Gauss product example. Top left) $\hat{I}_{\rm AME}$ as a function of $\Delta$, scaled by the true value.  Top right)  The actual error $\vert \hat{I}_{\rm AME}/I_{\rm true} -1\vert$ (red), the estimated uncertainty from the sample mean calculation (black) and the total estimated uncertainty (blue) as a function of $\Delta$.  Bottom left)  $\hat{I}_{\rm HME}$ scaled by the true value as a function of $\Delta$.    Bottom right) The actual error $\vert \hat{I}_{\rm HME}/I_{\rm true} -1\vert $ and the estimated uncertainty as a function of $\Delta$. The error bars in the left plots correspond to the estimated uncertainty. }
   \label{fig:evidence10D}
\end{figure}

The arithmetic mean calculation was performed at each of $500$ values of $\Delta$ as in the one-dimensional case, with $10^5$ samples in each AME run.  The results are shown in Fig.~\ref{fig:evidence10D}.  As is seen, for values of $\Delta$ around $1.5$, the uncertainty is about 1~\%.  The method does not show any systematic biases for $\Delta>1$, and the estimated uncertainty is again a good estimator for the error.  At small $\Delta$, where a small number of MCMC samples are used, the correlation between the MCMC samples produces some systematic errors in the evaluation of $I$.

The results for the HME estimator are also shown in Fig.~\ref{fig:evidence10D}.  We see that accuracies of a few tens of \% are achieved, but only in a narrow $\Delta$ range.  For $\Delta>2$, the error is more than $100$~\% and the HME estimate is no longer useful.  Also, the estimated uncertainty is too low and does not provide a reliable estimate of the true error.  The HME method is clearly already running into trouble at this level of complexity.

The target density is again a multivariate Gaussian, and the Laplace method works well, yielding 
$\hat{I}_{\rm L}/I_{\rm true}=0.977$.

\subsection{Gaussian Shell}
We now move beyond simple unimodal Gaussian type target densities and consider a function in $D$ dimensions with degenerate modes lying on a $D-1$ dimensional surface of fixed radius, a Gaussian shell:
\begin{equation}
\label{2shell}
f(\vec{\lambda}| \vec{c},r,\sigma)=\frac{1}{\sqrt{2\pi\sigma^2}}\exp\left(-\frac{(\vert \vec{\lambda}-\vec{c}\vert-r)^2}{2\sigma^2}\right)\; .
\end{equation}
This function is centered at $\vec{c}$ with degenerate modes along a surface of radius $r$.  The value of the function  decreases away from the modal surface along a radius according to a Gaussian shape with standard deviation $\sigma$.  The integral of this function can be evaluated using 
spherical coordinates centered at $\vec{c}$, where $\rho$ is the radial coordinate in the space, so that

$$I = \frac{1}{\sqrt{2\pi\sigma^2}} \int \exp\left(-\frac{(\rho-r)^2}{2\sigma^2}\right)\; dV $$

The volume element, integrated over the angular coordinates, is $dV = S_{D-1} \rho^{D-1} d\rho$ with 
$S_{D-1}=2\pi^{D/2}/\Gamma(D/2)$, so that we have 

$$I = \frac{ \sqrt{2}\pi^{(D-1)/2}}{\Gamma(D/2)\sigma}\int_0^{\rho_{max}} \rho^{D-1}\exp\left(-\frac{(\rho-r)^2}{2\sigma^2}\right) d\rho\; .$$

We are left with a one-dimensional integral that can be easily calculated numerically to high precision.  Note that we have assumed that the integral in the region outside $\rho_{\rm max}$ (the corners in the hypercube) is vanishingly small.  This is the case for the examples considered in this article.

 For the three examples below, we use the following settings: radius $r=5$, width $\sigma=2$ and $\vec{c}=\vec{0}$.  The integration region extends from $-10,+10$ in each dimension.

\subsubsection{2-Dimensional Gaussian-shell}
\label{cal2d}
\begin{figure}[htbp] 
   \centering
   \includegraphics[width=7cm]{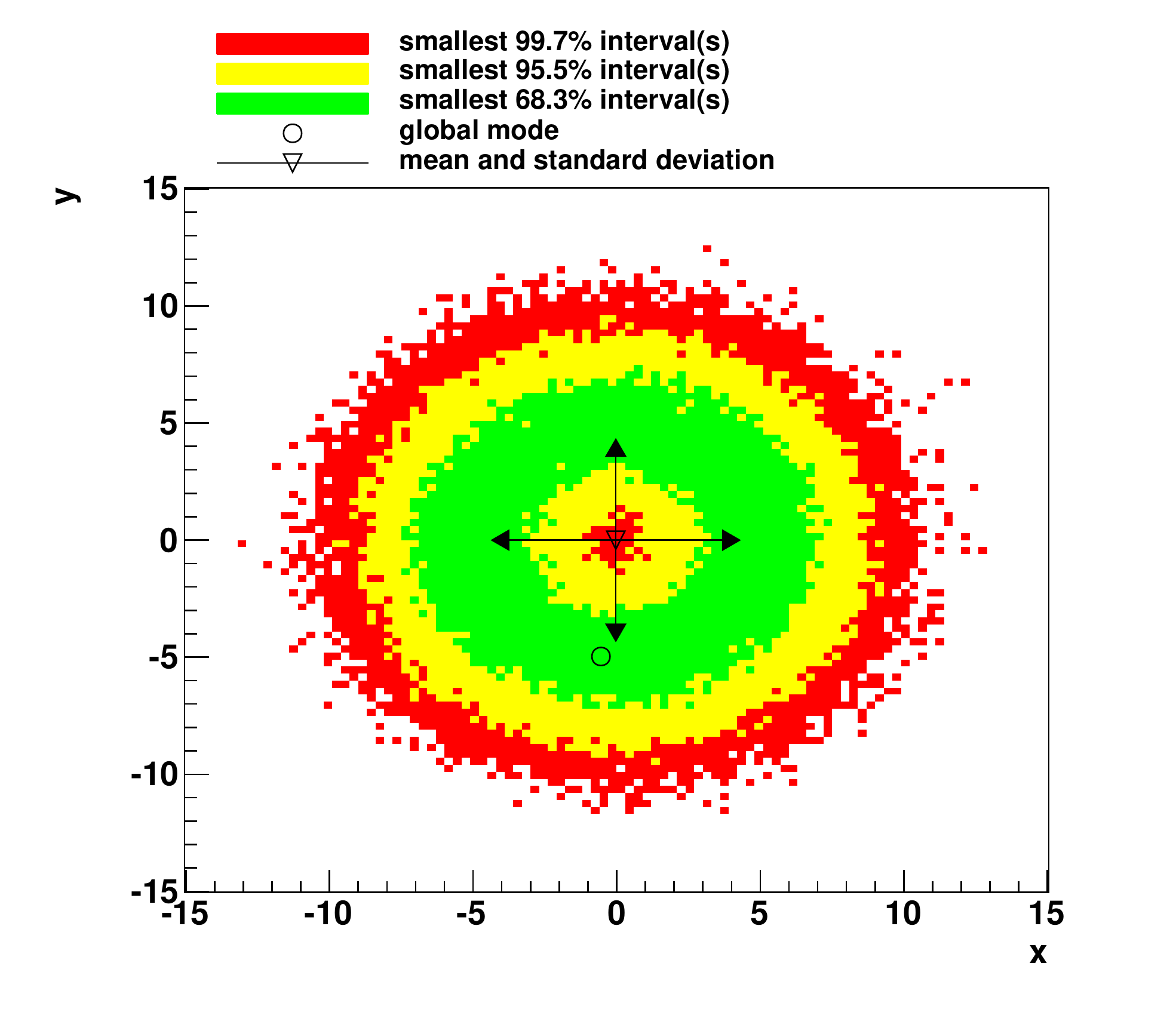}    \includegraphics[width=7cm]{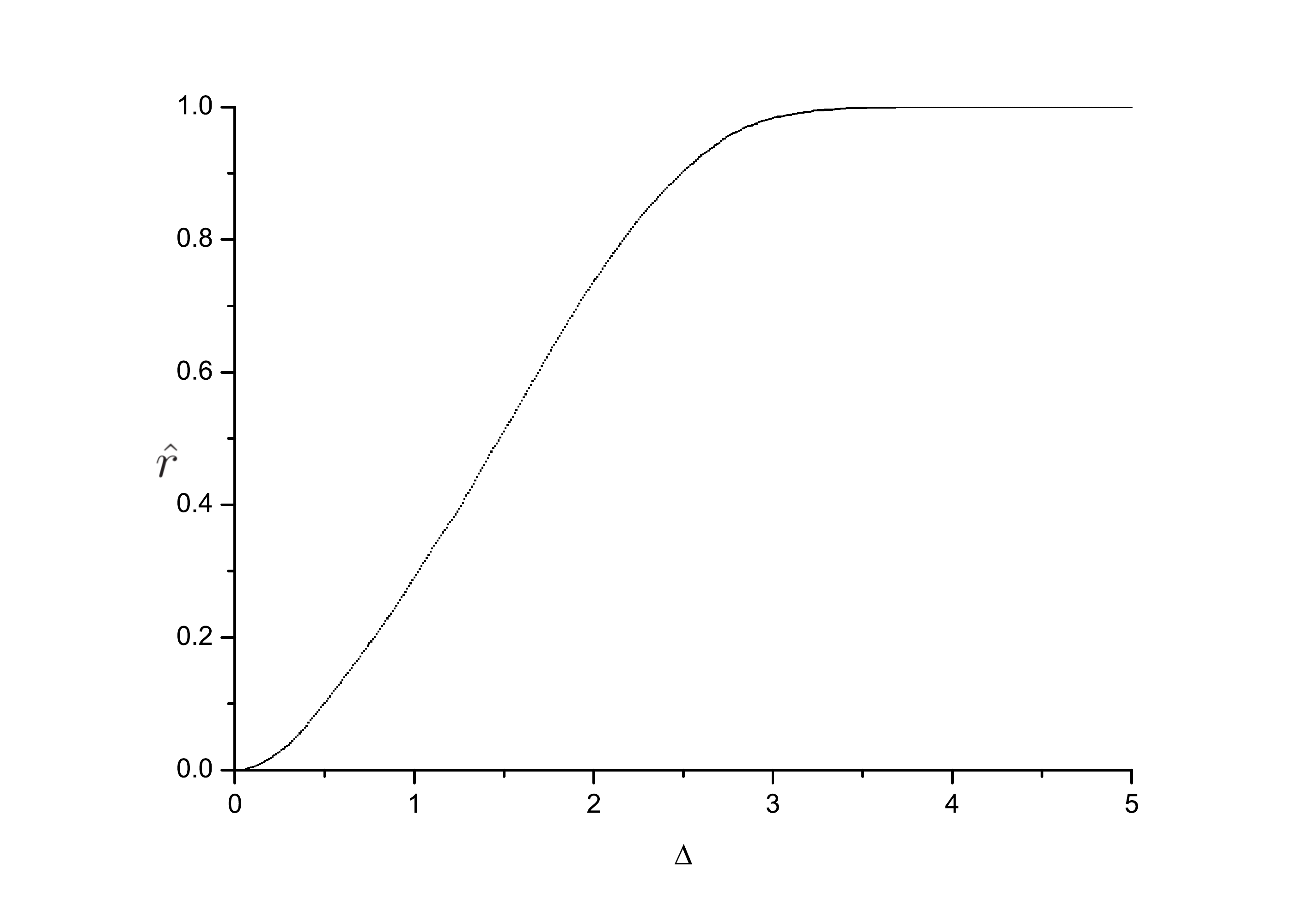} 
   \caption{Left) Distribution of samples from the MCMC algorithm BAT~\cite{ref:BAT} for the two-dimensional shell example.   The small circle indicates the location of the mode found from the MCMC samples. The arrows indicate the mean and $\pm1$ standard deviation ranges.  Right) Fraction of MCMC samples falling within the hypercube of side length $2\Delta$ as a function of the value of $\Delta$ (in units of the standard deviation of the distribution).}
   \label{fig:2DShell}
\end{figure}

 The parameter values result in $I_{\rm true} =7.86\cdot 10^{-4}$.  We use the BAT code to produce $10^5$ MCMC samples from the target density, yielding an effective sample size $N_{\rm ESS}=3.9 \cdot 10^4$. The sample distribution from the MCMC as well as the estimate of $\hat{r}$  as a function of $\Delta$ are shown in Fig.~\ref{fig:2DShell}.

As can be seen in the figure, the MCMC has produced a reasonable sample distribution.  The location of the mode from the posterior samples happens to be close to $(0,-5)$ and is indicated in the figure (note that $\vec{\lambda}=(x,y)$ in the figure).  The lack of a single mode is not a problem for the AME and HME algorithms, but we no longer expect the Laplace method to give sensible results.  The mean values of $(x,y)$ are very close to $(0,0)$ and the standard deviation in each direction is about $5$ units.  The hypercube centered at the mode found from the MCMC samples and with $\Delta=1$ contains about $20$~\% of the samples, and the hypercube with $\Delta=2$ contains about $75$~\% of the samples.

\begin{figure}[htbp] 
   \centering
   \includegraphics[width=7cm]{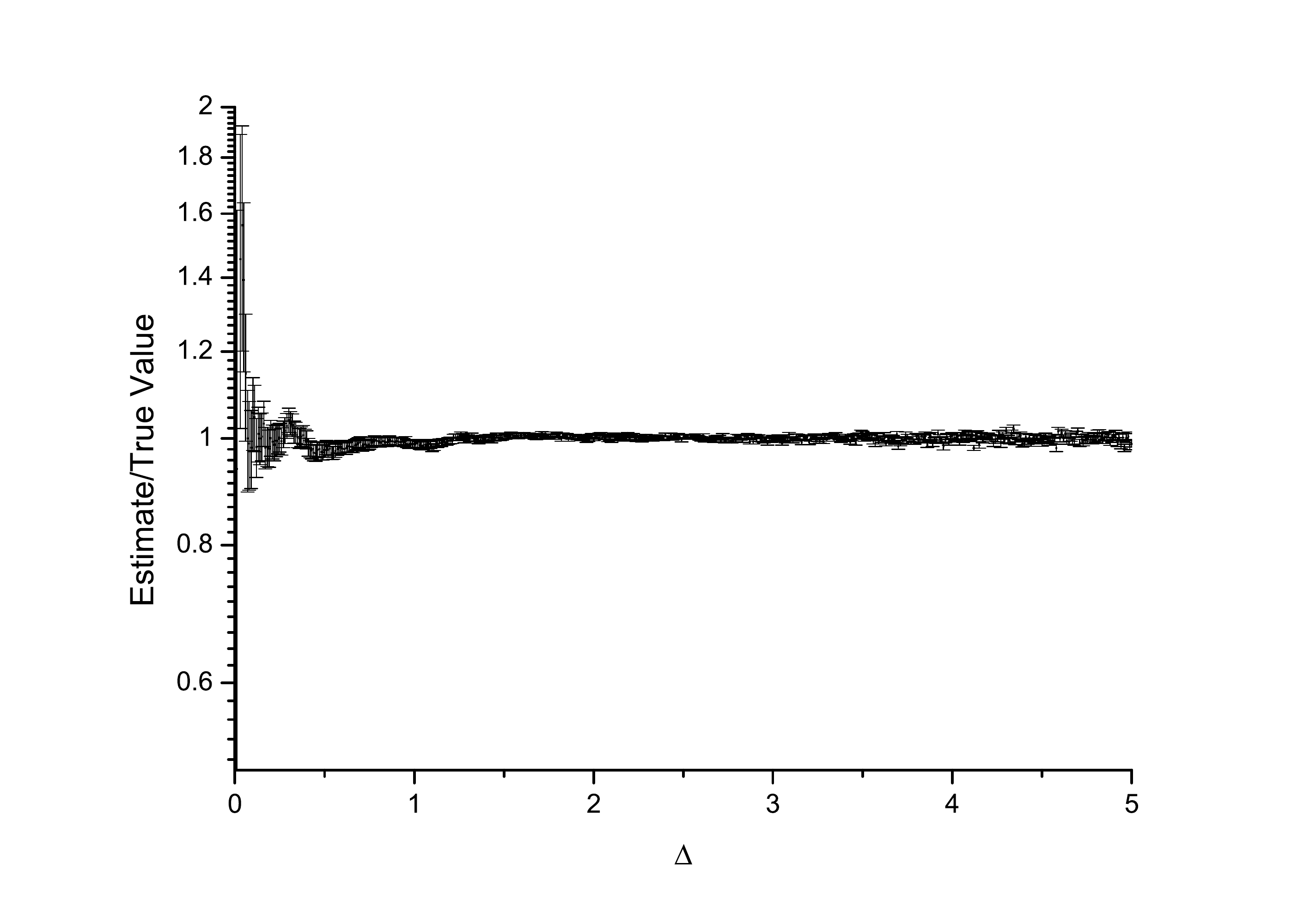}    \includegraphics[width=7cm]{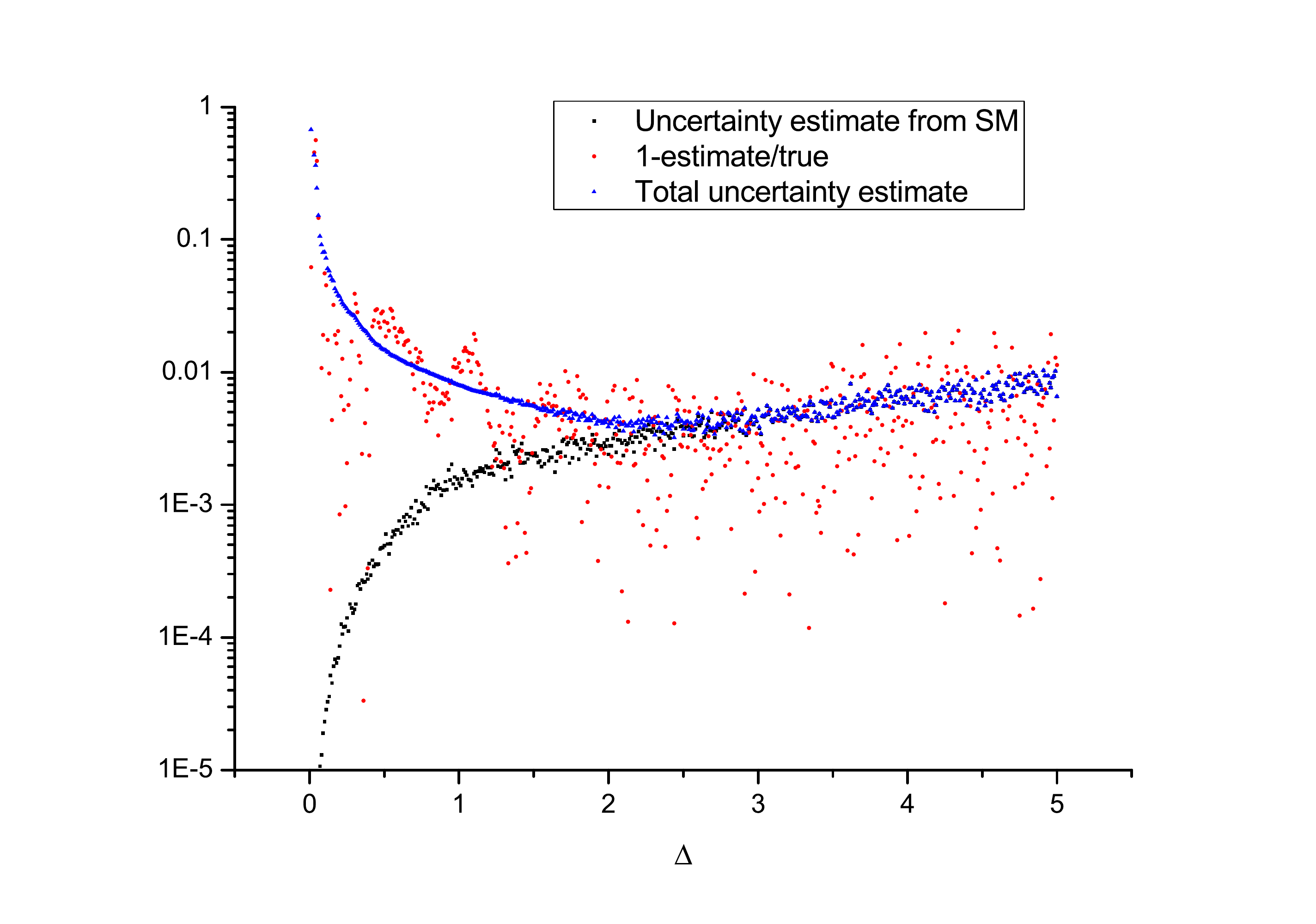} 
   \includegraphics[width=7cm]{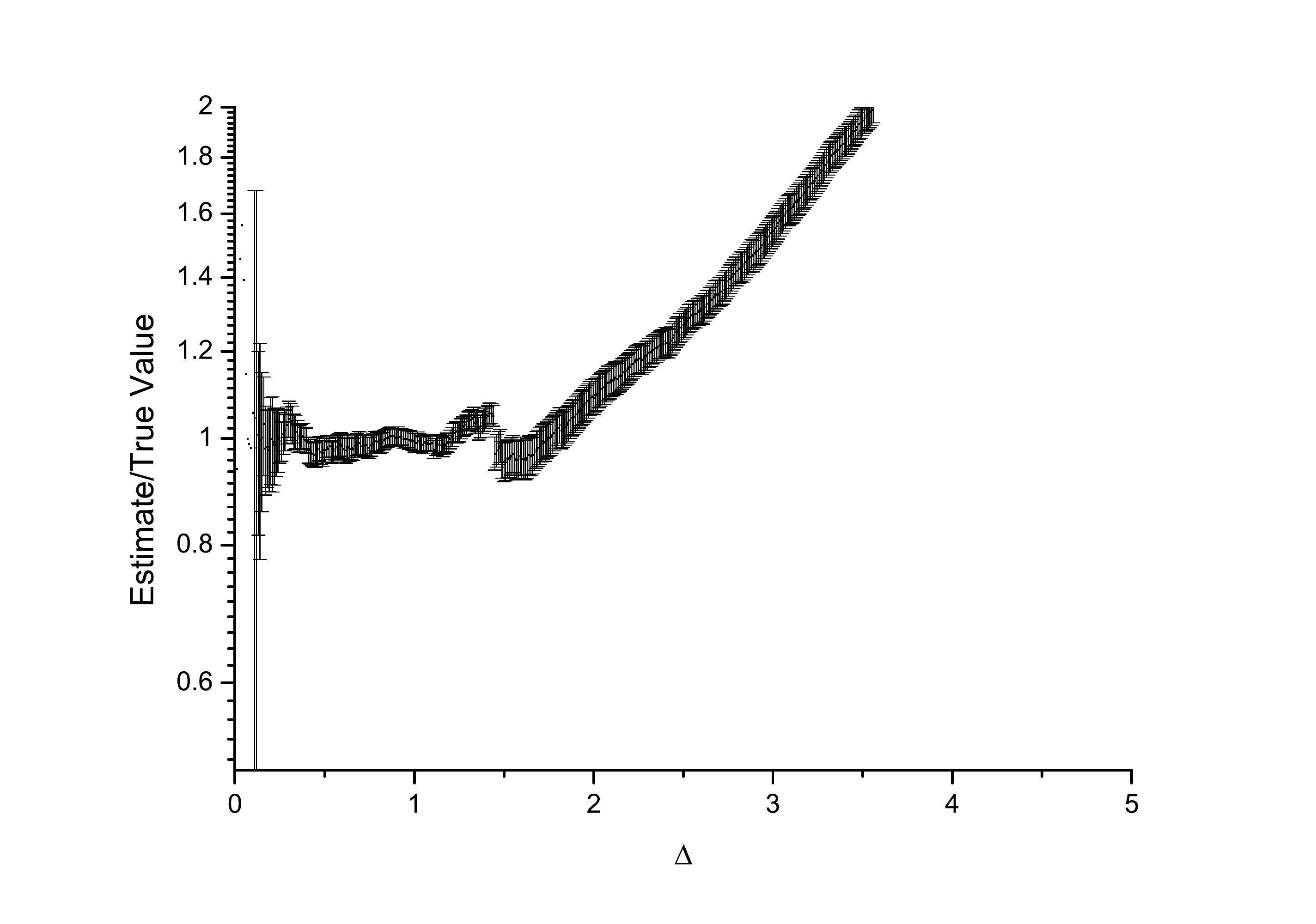}    \includegraphics[width=7cm]{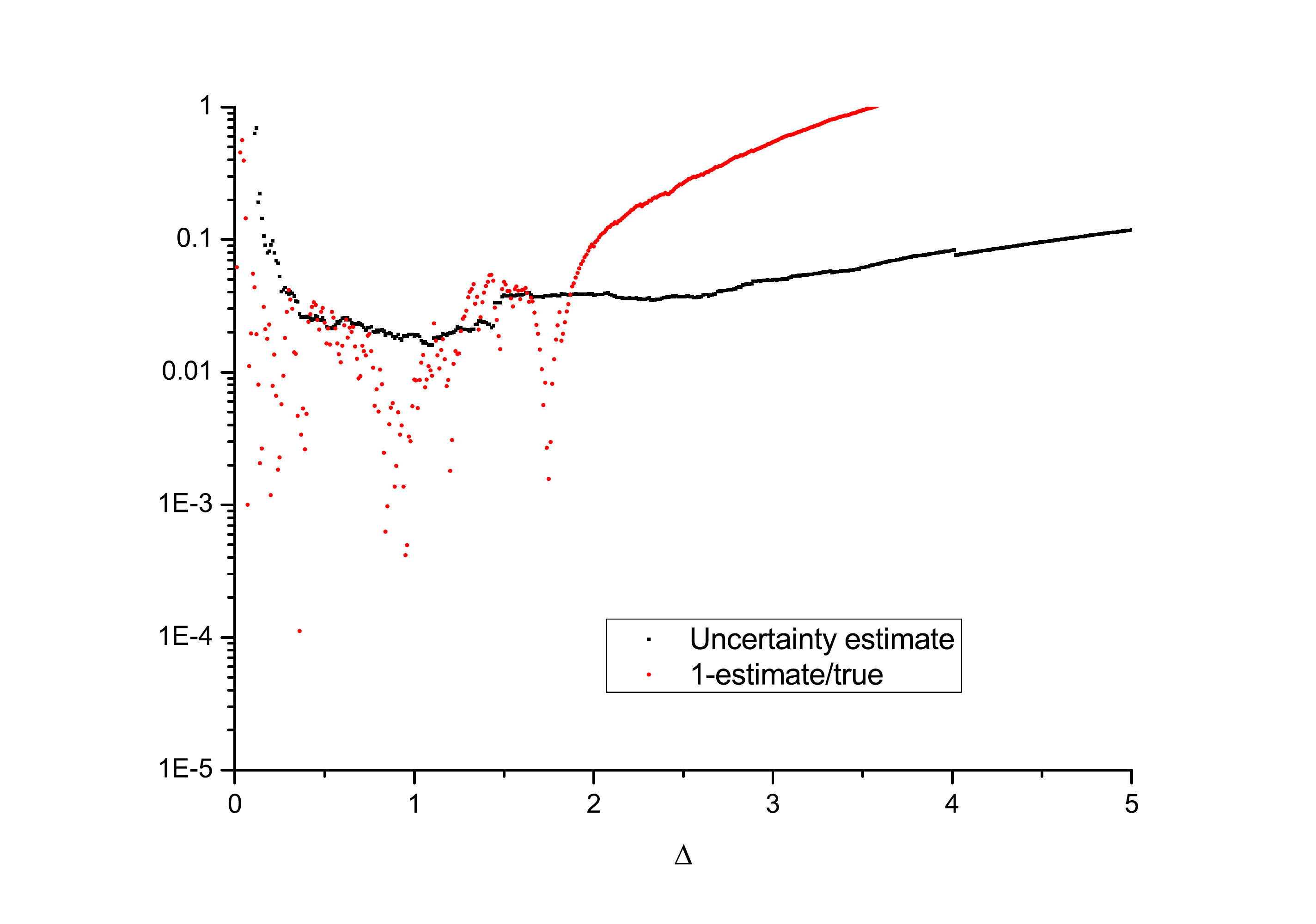} 
   \caption{Two-dimensional Gaussian shell example. Top left) $\hat{I}_{\rm AME}$ as a function of $\Delta$, scaled by the true value.  Top right)  The actual error $\vert \hat{I}_{\rm AME}/I_{\rm true} -1\vert$ (red), the estimated uncertainty from the sample mean calculation (black) and the total estimated uncertainty (blue) as a function of $\Delta$.  Bottom left)  $\hat{I}_{\rm HME}$ scaled by the true value as a function of $\Delta$.    Bottom right) The actual error $\vert \hat{I}_{\rm HME}/I_{\rm true} -1\vert $ and the estimated uncertainty as a function of $\Delta$. The error bars in the left plots correspond to the estimated uncertainty.}
   \label{fig:evidence2Dshell}
\end{figure}

We again use $10^5$ samples for our sample mean calculations at each of the values of $\Delta$.  The results for $\hat{I}_{\rm AME}$ are shown in the top plots in Fig.~\ref{fig:evidence2Dshell}, and we see that there is no difficulty in achieving a good result for the integral despite not having a simple mode for the target distribution.  The accuracy of the calculation is good, and the uncertainty is better than 1~\% for a wide range of $\Delta$, despite the rather small number of samples in the MCMC and AME calculations.  We again find that our estimated uncertainty gives a good reproduction of the actual error.

The HME evaluations are  also given in Figs.~\ref{fig:evidence2Dshell}.  Here we find good performance (few ~\% level accuracy) up to $\Delta=2$, at which point the HME calculation starts to systematically deviate from the correct value.  In this case, the estimated uncertainty does not give a reliable indication of the actual error for $\Delta>2$ and in fact the uncertainty is grossly underestimated.  This is a result of the missing MCMC samples at very small $f(\vec{\lambda})$.   The volume term in the numerator in Eq.~\ref{eq:HME} grows as $\Delta$ is increased, but is not properly compensated by large terms that should appear in the denominator from small values of  $f(\vec{\lambda})$.  The inability to diagnose this behavior implies that the HME is unreliable.

As expected, the Laplace method does not work for the Gaussian shell situation.  For the two-dimensional example considered here, $\hat{I}_{\rm L}/I_{\rm true}=0.03$.

\subsubsection{10-Dimensional Gaussian-shell}

\begin{figure}[htbp] 
   \centering
   \includegraphics[width=7cm]{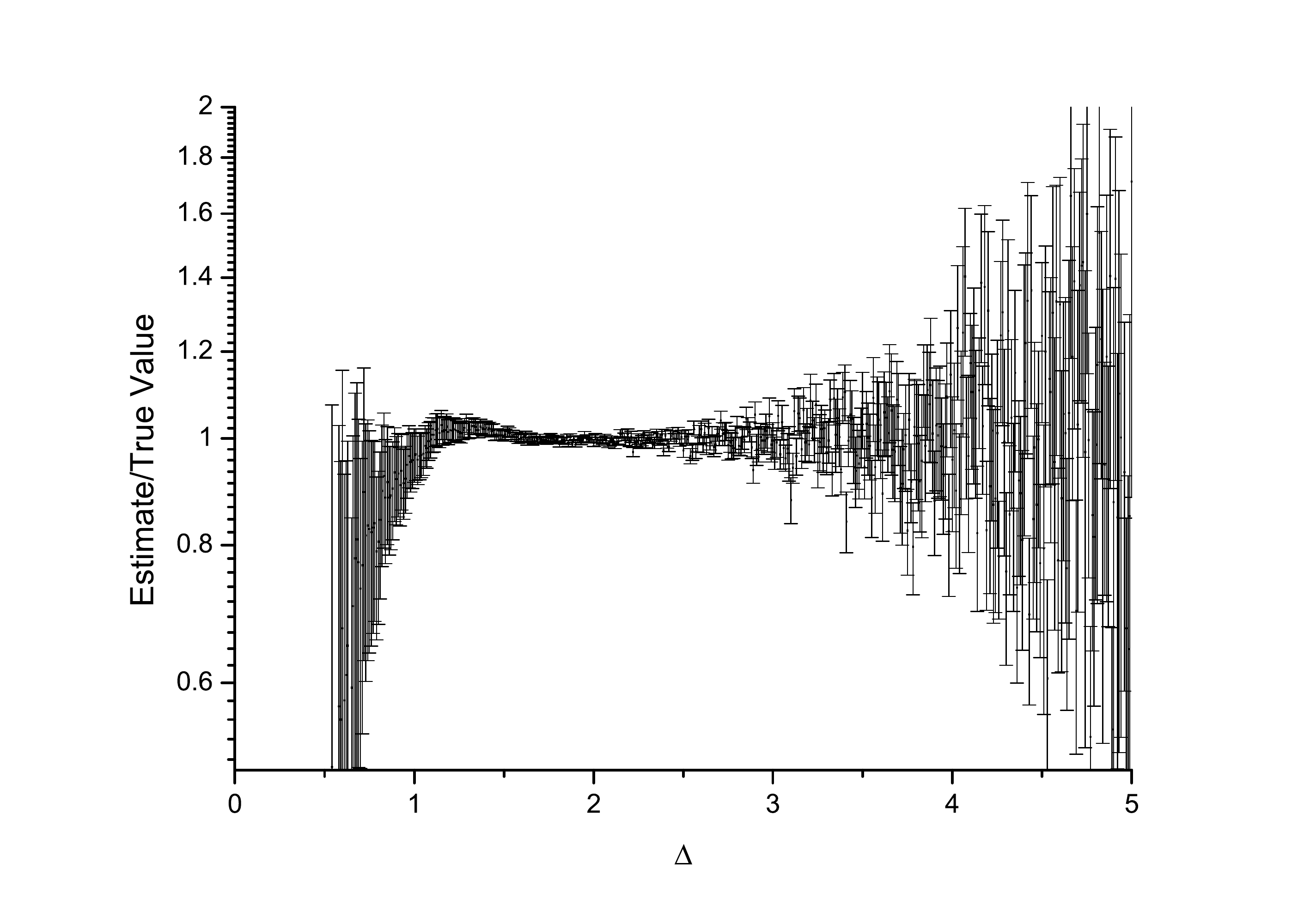}    \includegraphics[width=7cm]{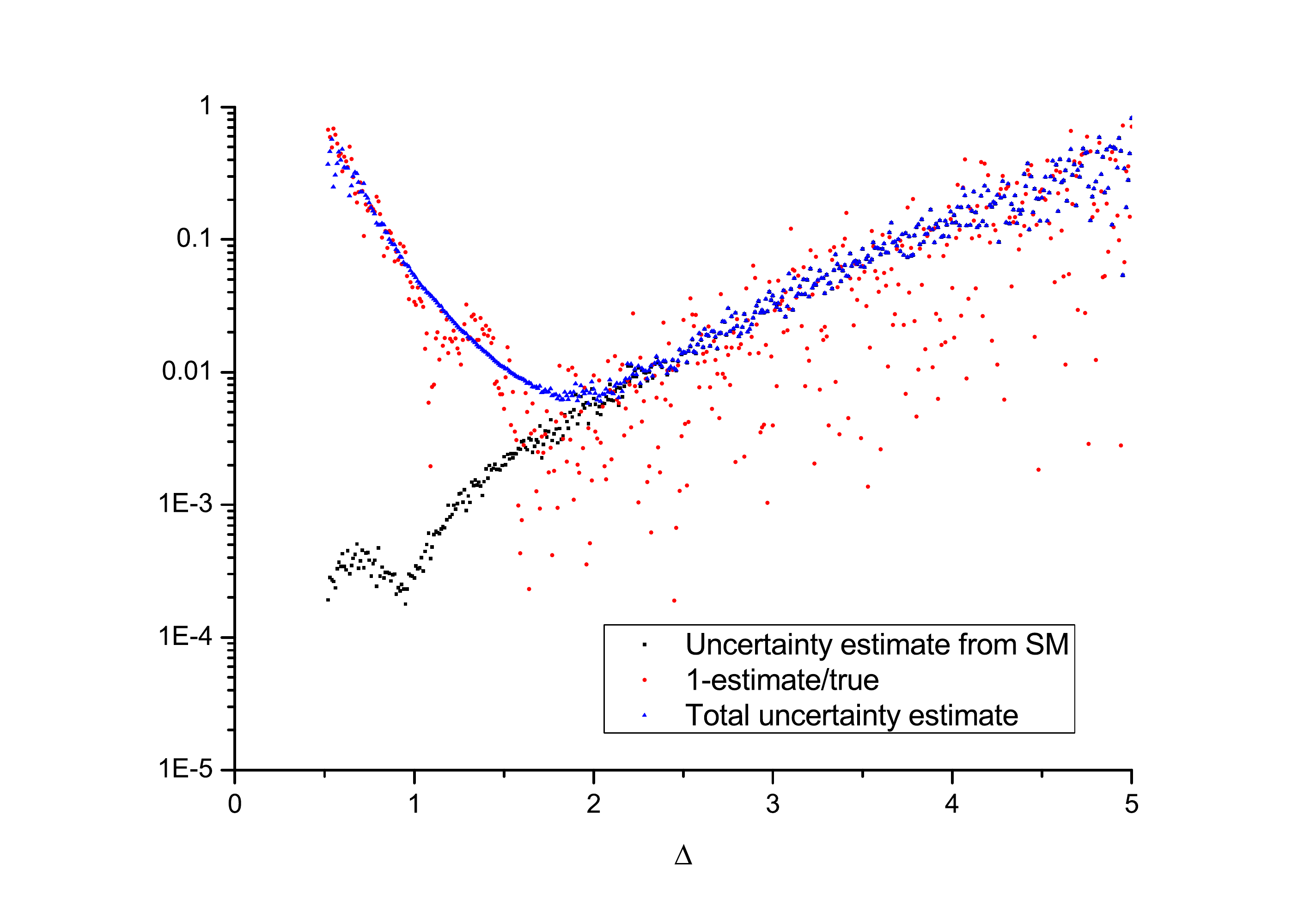} 
   \includegraphics[width=7cm]{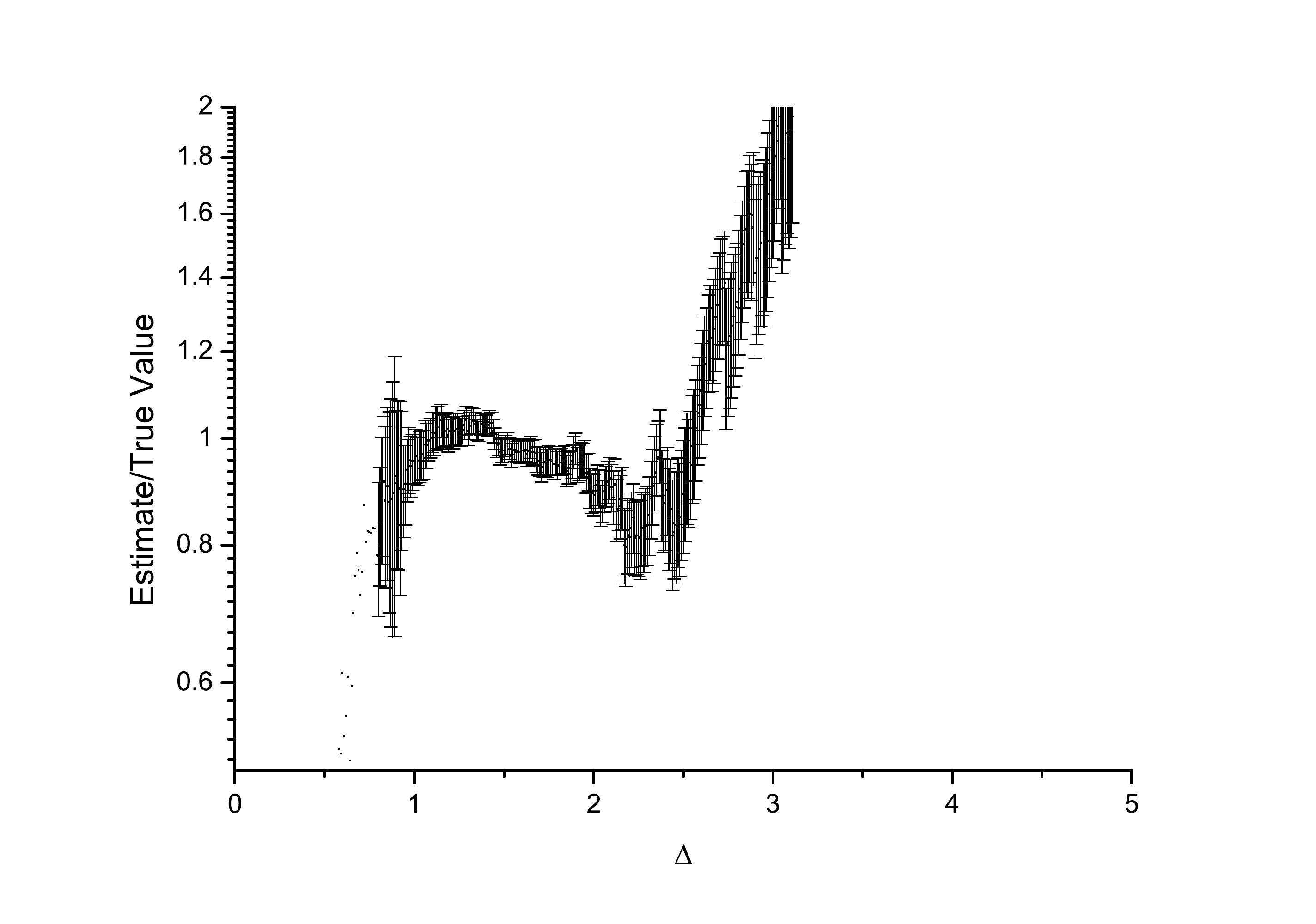}    \includegraphics[width=7cm]{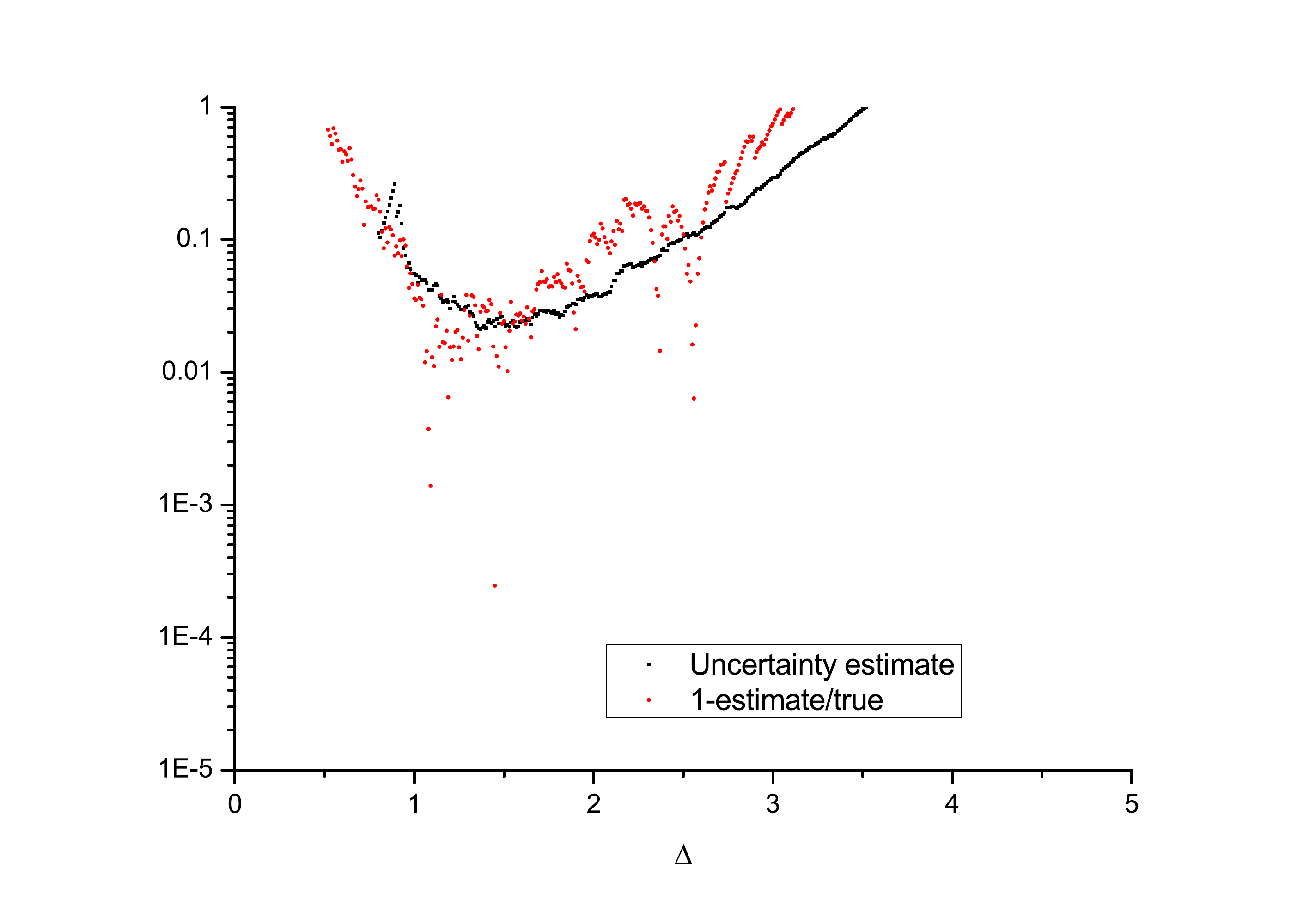} 
   \caption{Ten-dimensional Gaussian shell example. Top left) $\hat{I}_{\rm AME}$ as a function of $\Delta$, scaled by the true value.  Top right)  The actual error $\vert \hat{I}_{\rm AME}/I_{\rm true} -1\vert$ (red), the estimated uncertainty from the sample mean calculation (black) and the total estimated uncertainty (blue) as a function of $\Delta$.  Bottom left)  $\hat{I}_{\rm HME}$ scaled by the true value as a function of $\Delta$.    Bottom right) The actual error $\vert \hat{I}_{\rm HME}/I_{\rm true} -1\vert $ and the estimated uncertainty as a function of $\Delta$. The error bars in the left plots correspond to the estimated uncertainty.}
      \label{fig:evidence10Dshell}
\end{figure}

Here $\ln(I_{\rm true}) =-32.16$.  In a first calculation, we use the BAT code to produce $10^5$ MCMC samples from the target density, yielding an effective sample size $N_{\rm ESS}=3.9  \cdot 10^4$ and calculate the evidence. We again use $10^5$ samples for our sample mean calculations.  The results for the AME and HME evaluations are given in Fig.~\ref{fig:evidence10Dshell}.  For the arithmetic mean calculation, we see the same pattern as in the previous examples.  For small values of $\Delta$, the uncertainty coming from the small number of MCMC samples dominates.  However, sub~\% errors are possible for $\Delta \sim 1.5$, which corresponds to $\hat{r}\approx 0.2$.  As $\Delta$ increases, the uncertainties from the sample mean calculation dominate since we move to regions of the space that do not contain significant probability mass.  The estimated uncertainty is again accurate and can be used as a guide to choose the optimal value of $\Delta$ as we discuss below.

The HME estimate achieves few ~\% accuracy at a somewhat smaller value of $\Delta$ than the optimal for the sample mean calculation.  The estimated uncertainty is again tends too small at larger $\Delta$ and is not reliable.

As expected, the Laplace method does not work well and yields $\hat{I}_{\rm L}/I_{\rm true}=0.08$.

As a check that these results are not due to small MCMC sample size, the calculations were redone for $10^6$ MCMC samples.  The optimal value of $\Delta$ changes somewhat for the sample mean calculation, but otherwise all results are basically as before.  The systematic behavior of the $\vert \hat{I}_{\rm HME}/I_{\rm true} -1\vert $ is the same as for the smaller MCMC sample size; no significant improvement in performance was found with the 10 times large MCMC sample size.

\subsubsection{50-Dimensional Gaussian-shell}

\begin{figure}[htbp] 
   \centering
   \includegraphics[width=7cm]{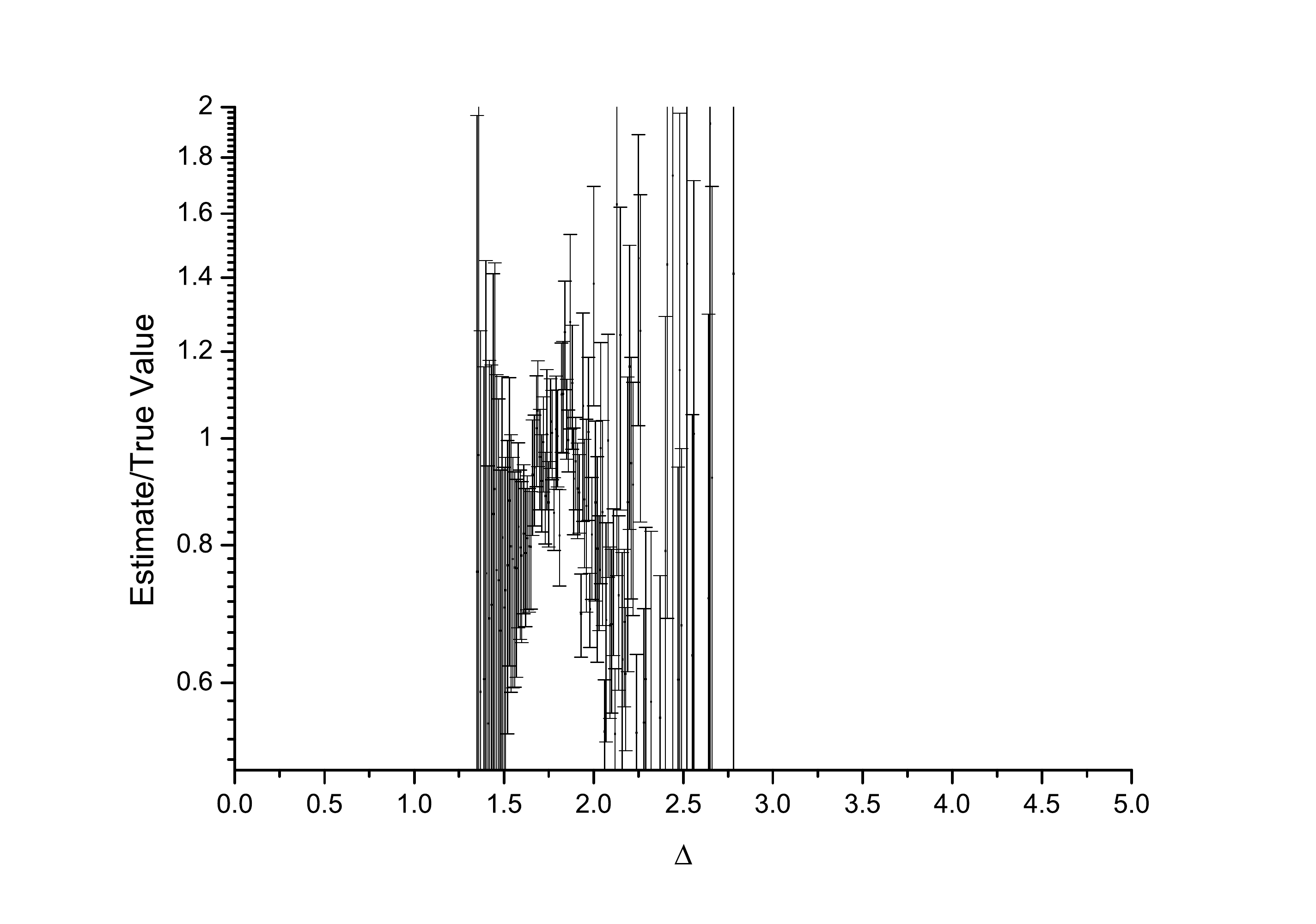}    \includegraphics[width=7cm]{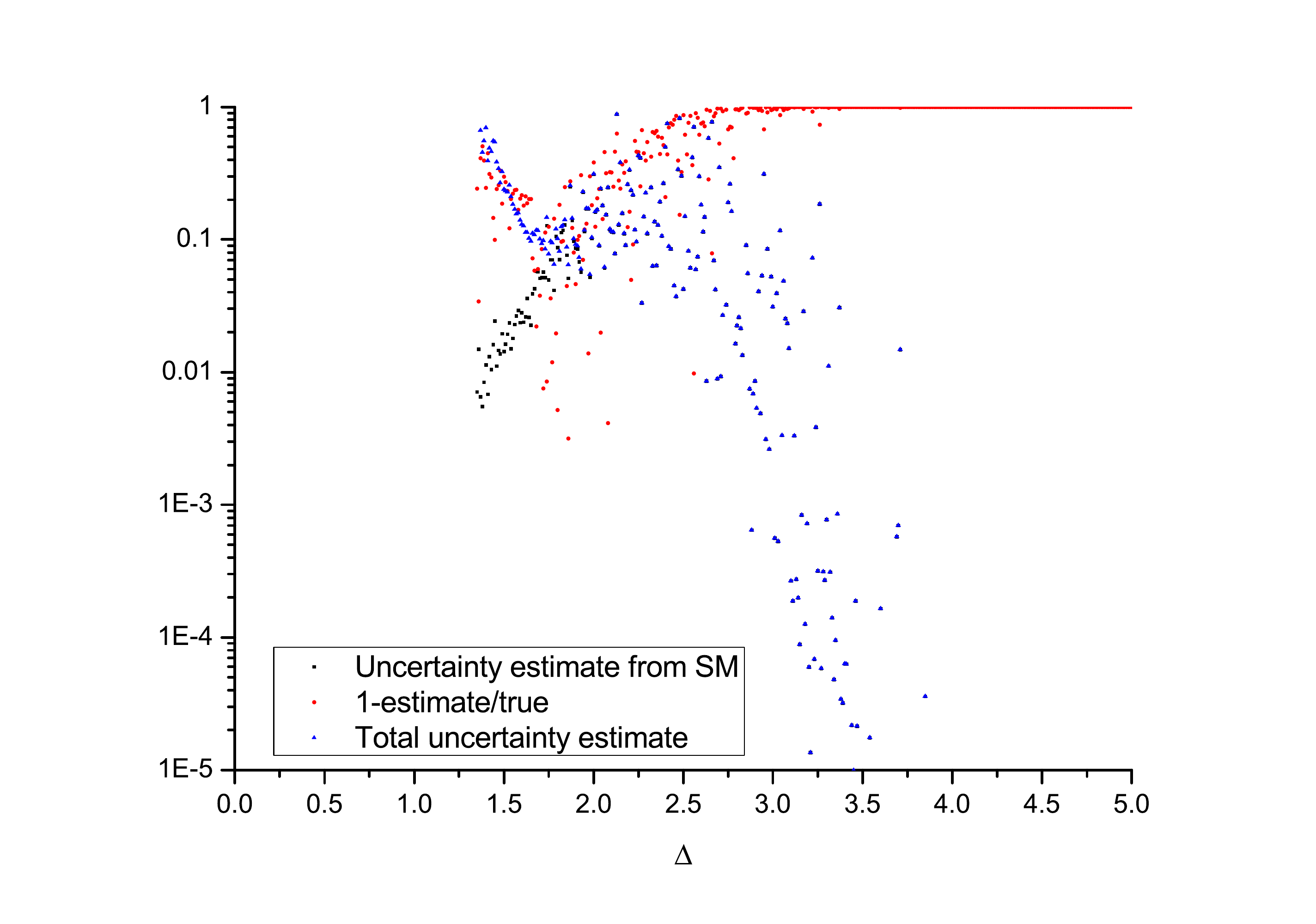} 
   \includegraphics[width=7cm]{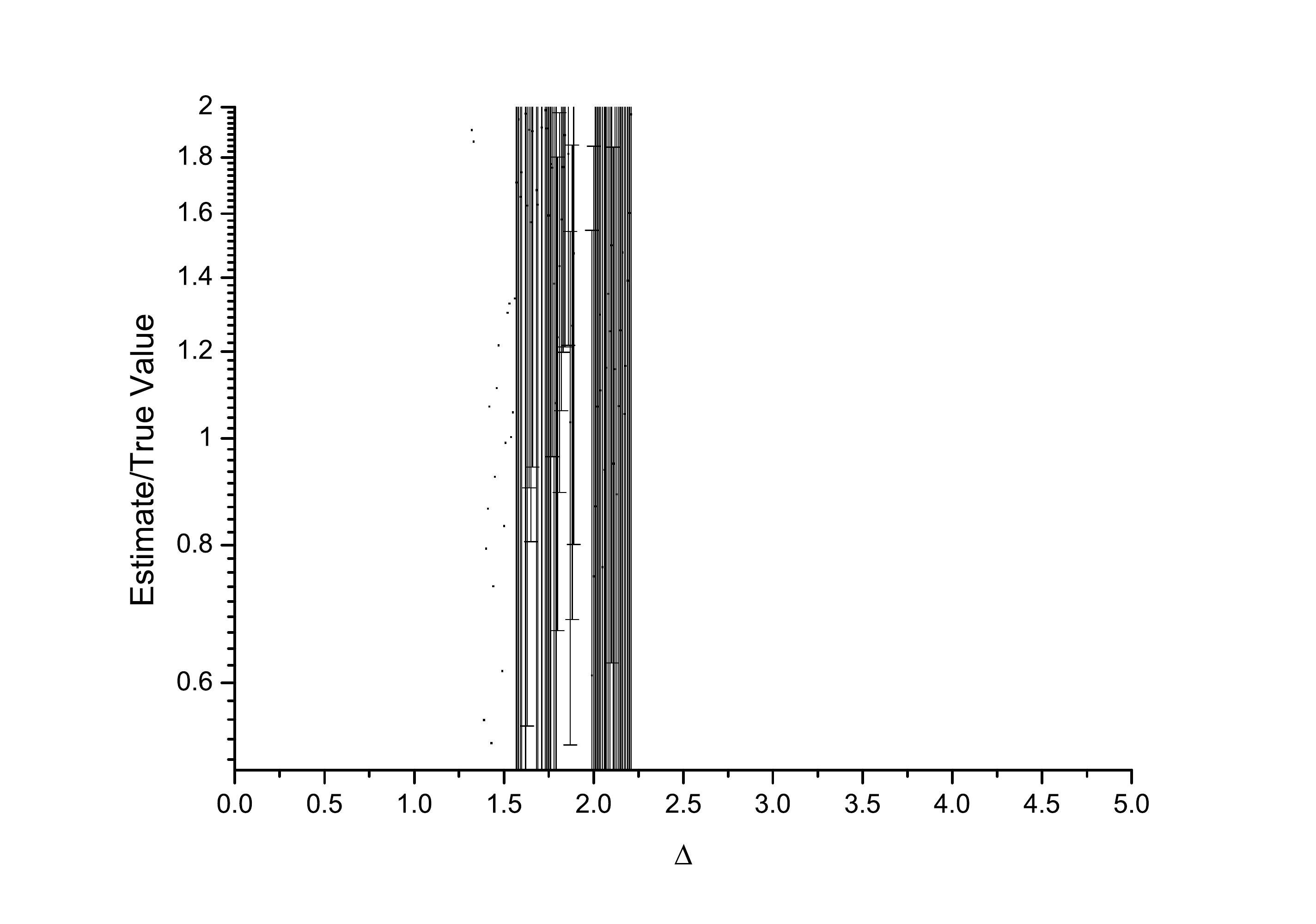}    \includegraphics[width=7cm]{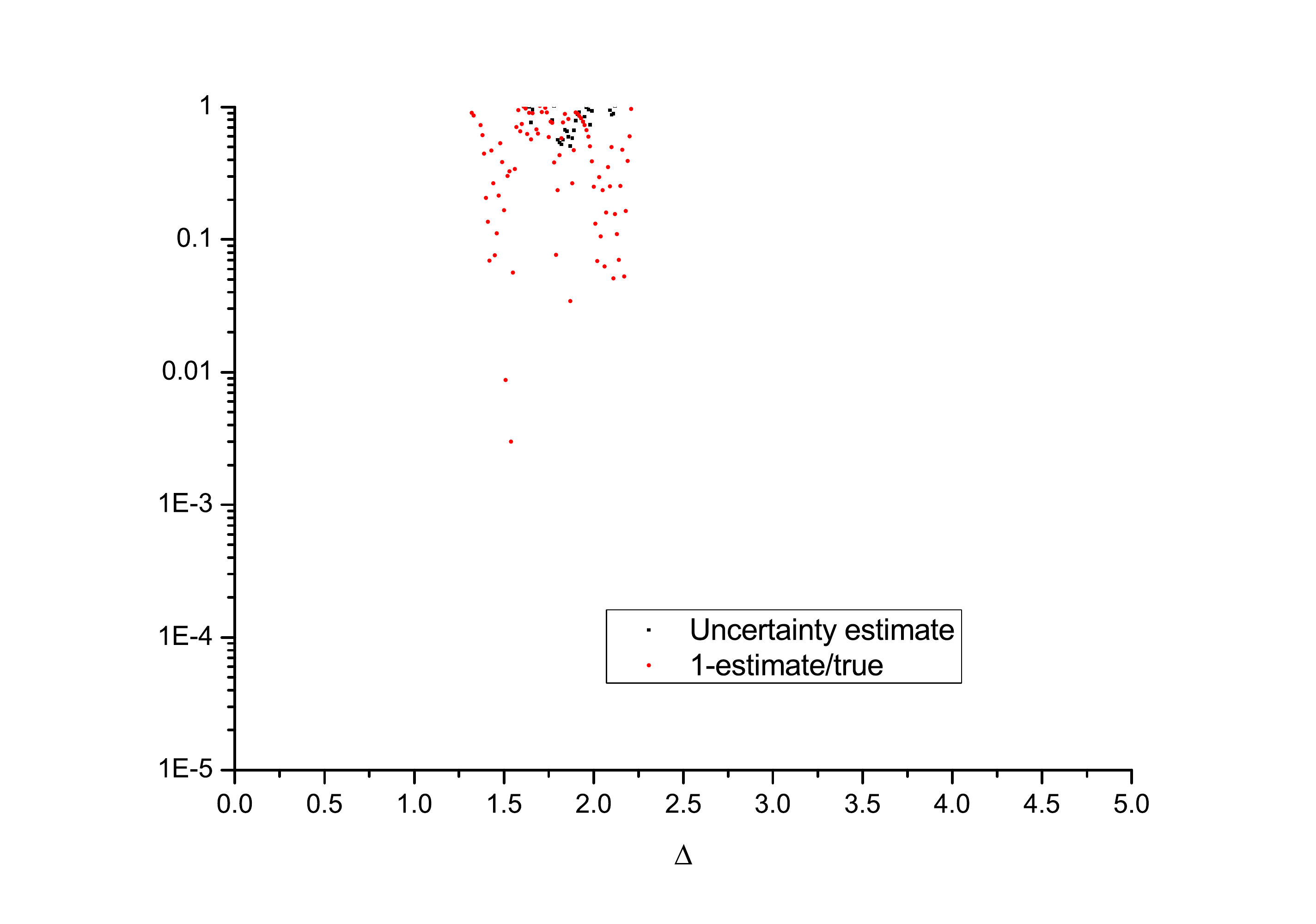} 
   \caption{50-dimensional Gaussian shell example. Top left) $\hat{I}_{\rm AME}$ as a function of $\Delta$, scaled by the true value.  Top right)  The actual error $\vert \hat{I}_{\rm AME}/I_{\rm true} -1\vert$ (red), the estimated uncertainty from the sample mean calculation (black) and the total estimated uncertainty (blue) as a function of $\Delta$.  Bottom left)  $\hat{I}_{\rm HME}$ scaled by the true value as a function of $\Delta$.    Bottom right) The actual error $\vert \hat{I}_{\rm HME}/I_{\rm true} -1\vert $ and the estimated uncertainty as a function of $\Delta$. The error bars in the left plots correspond to the estimated uncertainty. }
   \label{fig:evidence50Dshell}
\end{figure}

As an extreme example, we considered a 50-dimensional Gaussian shell.  Here the modal surface is a 49-dimensional hypersphere and $\ln(I_{\rm true}) =-169.8$.  The BAT code was used to initially produce $10^5$ MCMC samples from the target density, yielding an effective sample size $N_{\rm ESS}=3.9  \cdot 10^4$. The values of $\hat{r}$ increase rapidly from $\hat{r}\approx 0$ at $\Delta=1.8$ to $\hat{r}\approx 0.85$ at $\Delta=3$.  The standard deviations in each dimension is about $2$ units, so that $\Delta=5$ approximately covers the full support defined for the function.  The results for the AME and HME evaluations are given in Fig.~\ref{fig:evidence50Dshell}.  The best result for the sample mean calculation gives about $10$~\% accuracy, whereas the HME calculation is within $100$~\% of the correct result for a small range of $\Delta$ where $\hat{r}$ starts to increase.

We used $10^5$ samples for our sample mean calculations, although this is clearly too small a number for such a large dimensional volume.  The error from the sample mean calculation increases rapidly as we increase $\Delta$, and becomes completely unreliable for $\Delta>2$.  For such a large volume, the vast majority of sample mean evaluations are in regions where the target density is vanishingly small and the uncertainty grossly underestimates the true error.  In the next section, we discuss a choice of settings for the sample mean calculation and redo the calculation shown here.

As expected, the Laplace method does not work well and yields $\hat{I}_{\rm L}/I_{\rm true}=6 \cdot 10^{-4}$.

We again checked that these results are not due to small MCMC sample size, the calculations were redone for $10^7$ MCMC samples.  The optimal location of $\Delta$ changes to smaller values for the sample mean calculation and few~\% level accuracy is reached.  For the HME calculation, a small improvement is also observed, but otherwise all results are basically as before.

\section{Discussion}

Based on the results in the previous sections, we discuss now a procedure for choosing the value of $\Delta$ for both the sample mean and harmonic mean estimators.  As was seen in our examples, the uncertainty in the calculation for the AME estimator comes from two sources - the approximately binomial fluctuations in the number of MCMC samples included in our region of interest specified by $\Delta$, and the uncertainty coming from the sample mean calculation.  The first uncertainty can be estimated from the MCMC output, and can be used to define a value of $\hat{r}$ by specifying that this source of uncertainty should contribute half of the final uncertainty.  I.e., we find the value of $\hat{r}$ such that (see Eq.~\ref{eq:uncertainty})
$$\frac{\sigma_{r} }{\hat{r}} = \frac{\epsilon}{\sqrt{2}}$$
where $\epsilon$ is the target uncertainty.  We will use $\epsilon=0.01$ for our discussion below except for the fifty-dimensional Gaussian shell example, where we take $\epsilon=0.05$.  Once we have fixed $\hat{r}$ in this way, we then find the corresponding value of $\Delta$ and use this to calculate sample mean integrals with for a batch of samples, requiring a minimum of $10$ batches.  We use the variance of these $10$ calculations to determine how many batches will be needed to get the desired uncertainty; i.e.,
$$\frac{ \sigma_{I_{\Delta}}}{\hat{I}_{\Delta}}  = \frac{\epsilon}{\sqrt{2}} \; . $$
The results for the examples given in the previous sections using this procedure for fixing the parameters of the algorithm is given in Table~\ref{tab:summary}.  As is seen, the range of values for $\Delta$ is relatively narrow and only grows slowly with the complexity of the target function. The number of sample mean calculations however depends strongly on the complexity of the problem, and is also inversely dependent on the accuracy specified and on the size of the MCMC sample.  For a given specified accuracy, $\Delta$ is reduced as $N_{\rm MCMC}$ is increased, and this reduces the number of sample mean calculations necessary.  We find that the AME  algorithm gives a reliable estimate of the uncertainty for the examples chosen if the required number of sample mean calculations is not too large.  We conclude that the AME calculation of the integral of the target density using a reduced volume around the mode of the target works well for the types of cases we have studied.

\begin{table}
\scriptsize
\begin{tabular}{cccccrlcc}
\hline
 Test & $N_{\rm MCMC}$ & $N_{\rm ESS}$ & $\epsilon$ & $\Delta$  & $N_{\rm SM}$ & $I_{\rm true}$ & $1-\frac{\hat{I}_{\rm AME}}{I_{\rm true}}$  &$\sigma_{I_{\rm AME}}/I_{\rm AME}$ \\
 \hline
  1D Gaussian & $1 \cdot 10^5$&$3.9 \cdot 10^4$ & $0.01$ & $0.44$ &$18$ & $6.46\cdot 10^{-68}$&$0.011$ &$0.010$  \\
  \hline
  10D Gaussian & $1 \cdot 10^5$&$3.5 \cdot 10^4$ & $0.01$& $1.56$ &$652000$ & $9.09\cdot 10^{-124}$&$0.008$ &$0.010$  \\
   \hline
 2D shell & $1 \cdot 10^5$&$3.9 \cdot 10^4$ & $0.01$& $1.12$ &$6510$ & $7.71\cdot 10^{-4}$&$0.009$ &$0.010$  \\
 \hline
  10D shell & $1 \cdot 10^5$&$3.9 \cdot 10^4$ & $0.01$& $1.69$ & $18700$& $1.085\cdot 10^{-14}$&$0.009$ &$0.010$ \\
 \hline
  50D shell & $1 \cdot 10^5$&$3.9 \cdot 10^4$ & $0.05$& $1.89$ & $1352400$ & $1.81\cdot 10^{-74}$&$0.056$ &$0.046$   \\
 \hline
  10D shell & $1 \cdot 10^6$& $8.9 \cdot 10^5 $ & $0.01$& $1.13$ & $718$& $1.085\cdot 10^{-14}$&$0.010$ &$0.010$  \\
 \hline
  50D shell & $1 \cdot 10^7$ & $8.6 \cdot 10^6$ & $0.05$& $1.47$ & $19200$ & $1.81\cdot 10^{-74}$&$0.090$ &$0.044$   \\
 \hline
\end{tabular}
\caption{Summary of the results on different target functions for the AME  estimator of the normalizing integral.  $N_{\rm MCMC}$ is the number of posterior samples from the MCMC, $N_{\rm ESS}$ is the effective sample size, $\epsilon$ is the specified accuracy for the integral calculation, $\Delta$ is the multiplier of the standard deviation along each dimension chosen by the algorithm, $N_{\rm SM}$ is the number of samplings of the function used in the sample mean calculation, $I_{\rm true}$ is the true value of the integral, $1-\frac{\hat{I}_{\rm AME}}{I_{\rm true}}$ is the fractional error made in the calculation and $\sigma_{I_{\rm AME}}/I_{\rm AME}$ is the estimated fractional uncertainty from the calculation.}
\label{tab:summary}
\normalsize
\end{table}

For the HME estimator, the post-convergence samples of the MCMC are used in the calculation. For want of a better method, we fix $\Delta$ by requiring that $\hat{r}=0.5$ as this value was typically near optimal for the examples studied.  The results for the examples given above using this fixing of the algorithm is given in Table~\ref{tab:summary2}.

\begin{table}
\scriptsize
\begin{tabular}{cccclcc}
\hline
 Test & $N_{\rm MCMC}$ & $N_{\rm ESS}$  & $\Delta$ & $I_{\rm true}$ & $1-\frac{\hat{I}_{\rm HME}}{I_{\rm true}}$  &$\sigma_{I_{\rm HME}}/I_{\rm HME}$ \\
 \hline
  1D Gaussian & $1 \cdot 10^5$&$3.9 \cdot 10^4$  & $0.66$ & $6.46\cdot 10^{-68}$&$0.008$ &$0.010$  \\
  \hline
 10D Gaussian & $1 \cdot 10^5$&$3.5 \cdot 10^4$  & $1.74$ & $9.09\cdot 10^{-124}$&$1.129$ &$0.353$  \\
   \hline
  2D shell & $1 \cdot 10^5$&$3.9 \cdot 10^4$  & $1.47$ & $7.71\cdot 10^{-4}$&$0.019$ &$0.034$  \\
 \hline
  10D shell & $1 \cdot 10^5$&$3.9 \cdot 10^4$ & $1.85$ & $1.085\cdot 10^{-14}$&$0.066$ &$0.031$ \\
 \hline
  50D shell & $1 \cdot 10^5$&$3.9 \cdot 10^4$  & $2.46$ & $1.81\cdot 10^{-74}$&$23.251$ &$17.482$   \\
 \hline
  10D shell & $1 \cdot 10^6$&$8.9 \cdot 10^5 $ & $1.85$ & $1.085\cdot 10^{-14}$&$0.005$ &$0.004$  \\
 \hline
  50D shell & $1 \cdot 10^7$ &$8.6 \cdot 10^6 $  & $2.46$  & $1.81\cdot 10^{-74}$&$2.680$ &$0.707$   \\
 \hline
\end{tabular}
\caption{Summary of the results on different target functions for the HME  estimator of the normalizing integral with $\hat{r}$ set to $0.5$.  $N_{\rm MCMC}$ is the number of posterior samples from the MCMC, $N_{\rm ESS}$ is the effective sample size, $\Delta$ is the multiplier of the standard deviation along each dimension chosen by the algorithm, $I_{\rm true}$ is the true value of the integral, $1-\frac{\hat{I}_{\rm HME}}{I_{\rm true}}$ is the fractional error made in the calculation and $\sigma_{I_{\rm HME}}/I_{\rm HME}$ is the estimated fractional uncertainty from the calculation.}
\label{tab:summary2}
\normalsize
\end{table}

As can be seen from the table, and as discussed earlier, the HME calculation works well for the simple target functions considered, but does not produce good results for the more complicated target functions.  In particular, the estimated uncertainty does not provide a good estimate of the actual error, so that it is not possible to diagnose that the calcuclation is not performing well.  We therefore do not recommend the use of the HME estimator to calculate the normalization integral for anything but the simplest low-dimensional target densities,

The Laplace estimation works well in cases where the target density is well approximated by a (multivariate) Gaussian distribution.  If this is known to be the case, then this approximation is easily calculated and can be used.  However, it should be avoided if the shape of the target distribution is not well known. 

\section{Summary}
We have investigated techniques for the integration of the target density in cases where a MCMC algorithm has successfully run.  We do not attempt to modify the sampling of the target density, but only to provide a post-processor for an MCMC algorithm. From the MCMC, we have an estimate of the global mode and also the variance of the samples marginalized along each parameter dimension.  We use this information to define a hypercube centered on the global model and having side lengths proportional to the standard deviation along these directions, and then calculate the integral of the target function in the reduced volume using either an arithmetic mean or harmonic mean approach.  The fraction of MCMC samples within the reduced volume was used to estimate the integral of the target density over the full volume of interest.  This technique was tried on a variety of examples and also compared to a Laplace estimator. The key elements of the methods studied are:
\begin{itemize}
\item Given the MCMC has been run successfully, the evaluation of the normalization of the target function can be performed using any sub support of the support of the target function;
\item From the MCMC, we can find a point near the maximum of the target function, and we can perform the integration in a region which is in some ways optimal by centering the sub support on this point;
\item It is possible to also calculate an estimated accuracy for the integral.
\end{itemize}

Our conclusions are that the arithmetic mean calculation performed in a hypercube centered on the observed mode works well and provides a technique for calculating the normalization of the target density with a reliable uncertainty estimate.  On the other hand, the harmonic mean estimator only works well in situations where the range of values from the target density does not vary too widely, and the Laplace estimator is restricted for use on Gaussian shaped target distributions.

\section*{Acknowledgments}
The authors would like to thank Frederik Beaujean, Daniel Greenwald, Stephan Jahn and Kevin Kr\"oninger for many fruitful discussions.

\end{document}